\documentclass[showpacs,preprintnumbers,amsmath,amssymb,prb]{revtex4}

\usepackage{graphicx}
\usepackage{dcolumn}
\usepackage{bm}
\usepackage{amsfonts}
\usepackage{amssymb}

\newcommand{\no}[1]{}

\def\drawline#1#2{\raise 2.5pt\vbox{\hrule width #1pt height #2pt}}

\def\trian{\raise 1.25pt\hbox{$\scriptscriptstyle\triangle$}\nobreak\ }

\def\square{${\vcenter{\hrule height .4pt
        \hbox{\vrule width .4pt height 3pt \kern 3pt
        \vrule width .4pt}
        \hrule height .4pt}}$\nobreak\ }

\def\plus{\raise 1.25pt \hbox{$\scriptscriptstyle +$}\nobreak\ }

\begin{document}

\title{Boundary layer structure in turbulent Rayleigh-B\'{e}nard convection}
\author{Nan Shi, Mohammad S. Emran and J\"org Schumacher}
\affiliation{Institut f\"ur Thermo- und Fluiddynamik, \\
                 Technische Universit\"at Ilmenau, \\
                  Postfach 100565, D-98693 Ilmenau, Germany}
\date{\today}                  

\begin{abstract}
The structure of the boundary layers in turbulent Rayleigh-B\'{e}nard convection is 
studied by means of three-dimensional direct numerical simulations. We consider convection in a 
cylindrical cell at an aspect ratio one for Rayleigh numbers of $Ra=3\times 10^9$ and 
$3\times 10^{10}$  at fixed Prandtl number $Pr=0.7$. Similar to the experimental results in the same setup and for the same Prandtl number, 
the structure of the laminar boundary layers of the velocity and temperature fields is found 
to deviate from the prediction of the Prandtl-Blasius-Pohlhausen theory. Deviations decrease 
when a dynamical rescaling of the data with an instantaneously defined boundary 
layer thickness is performed and the analysis plane is aligned with the instantaneous 
direction of the large-scale circulation in the closed cell. Our numerical results demonstrate 
that important assumptions which enter existing classical laminar boundary layer theories 
for forced and natural convection are violated, such as the strict two-dimensionality of the 
dynamics or the steadiness of the fluid motion. The boundary layer dynamics consists of two
essential local dynamical building blocks, a plume detachment and a post-plume phase. The former is associated with 
larger variations of the instantaneous thickness of velocity and temperature boundary layer and a fully
three-dimensional local flow. The post-plume dynamics is connected with the large-scale circulation in the cell that penetrates the 
boundary region from above. The mean turbulence 
profiles taken in localized sections of the boundary layer for both dynamical phases are
also compared with solutions of perturbation expansions of the boundary layer equations 
of forced or natural convection towards mixed convection.  Our analysis of both boundary layers shows 
that the near-wall dynamics combines elements of forced Blasius-type and natural convection. 
\end{abstract}

\pacs{}
\maketitle
\section{Introduction}   
Turbulent Rayleigh-B\'{e}nard  convection can be initiated in a fluid which is confined between a cold isothermal plate at the top and a hot isothermal plate at the bottom, given a sufficiently strong temperature difference is sustained. In the turbulent regime, the majority of the heat is carried by convective transport through the layer or cell. Only in the vicinity of the top and bottom plates, where the fluid velocities are small, 
conductive transport takes over and becomes important. As in all other wall-bounded flows, boundary layers form. In the present system these are boundary layers of the velocity and temperature fields. The structure of these boundary layers turns out to be crucial for a deeper understanding of the local and global transport processes as discussed for example in a recent review (Ahlers {\it et al.} 2009). Furthermore, the boundary layers interact with a so-called 
large-scale circulation (LSC) that is always established in a closed turbulent convection cell. 
This LSC can take the form of a single roll for aspect ratios of order unity or multiple roll patterns for larger ones (du Puits et al. 2007a, van Reeuwijk et al. 2008, Bailon-Cuba et al. 2010, Mishra et al. 2011).  On the one hand, the LSC is triggered by packets of thermal plumes -- fragments of the thermal boundary layers which detach randomly from the top and bottom plates into the bulk of the cell. On the other hand, the fully established LSC with its complex three-dimensional dynamics can be expected to affect and partly even drive the laminar flow dynamics close to the walls. This interplay has not yet been studied in detail for cylindrical convection cells and provides one central motivation for the present work.  

From a global perspective the heat transport in a turbulent convection cell, which is measured by the dimensionless Nusselt number $Nu$, is a function of the three dimensionless control parameters in  Rayleigh-B\'{e}nard convection, namely the Rayleigh number $Ra$, the Prandtl number $Pr$ and the  aspect ratio $\Gamma$ of the convection cell, i.e. $Nu=f(Ra, Pr, \Gamma)$. Two scaling theories yield 
different predictions for the turbulent heat transport in convection based  on different assumptions 
on the boundary layer structure. While the scaling theory of Shraiman \& Siggia (Siggia 1994) is based on a turbulent boundary layer with a logarithmic profile for the mean streamwise velocity, Grossmann \& Lohse (2000) assume laminar boundary layers of  Prandtl-Blasius-Pohlhausen type (Prandtl 1905; Blasius 1908; Pohlhausen 1921) in order to estimate the boundary layer contributions to the thermal 
and kinetic energy dissipation rates. Such laminar boundary layer evolves in purely {\em forced convection}, i.e. for a laminar flow over a flat plate. The temperature is treated as a passive scalar (Pohlhausen 1921).

Measuring the boundary layer structure is, however, difficult in laboratory experiments 
for high-Rayleigh-number convection. The reason is that the thickness of the thermal boundary layer, 
$\delta_T$, decreases as the Rayleigh and thus the Nusselt number grow. This thickness is given by 
\begin{equation}
\delta_T=\frac{H}{2 Nu}\,,
\end{equation}
where $H$ is the height of the convection cell. For a convection flow at $Pr\sim {\cal O}(1)$, the corresponding velocity boundary layer will have a similar thickness of $\delta_v\sim \delta_T$ and 
will thus decrease similarly with increasing Rayleigh number (see e.g. Shishkina et al. 2010).  
Detailed measurements of boundary layer profiles at higher Rayleigh numbers ($Ra > 10^9$) 
require thus large devices such as the Barrel of Ilmenau for the convection in air (du Puits et al. 
2007; du Puits et al. 2010) or high-resolution particle image velocimetry, as possible for convection in water (Sun et al. 2008; Zhou \& Xia 2010). Statistical time-series analyses of the mean temperature 
and velocity profiles in the boundary layer yielded deviations from the predicted laminar 
Blasius profiles (du Puits et al. 2007; Zhou \& Xia, 2010). A dynamic rescaling of the data with respect to an instantaneous boundary layer thickness (which will be explained further below in the text) tends to bring it closer to the Blasius prediction in the water experiment by 
Zhou \& Xia (2010). The latter result was also confirmed by a series of two-dimensional direct 
numerical simulations by Zhou et al. (2010, 2011). However in both cases, the large-scale circulation is a (quasi-) two-dimensional flow which cannot fluctuate in the third direction.

Du Puits et al. (2007) concluded from their work that the deviations from the Blasius shape
arise due to the characteristic near-wall coherent structures -- so-called thermal plumes -- which permanently detach from the thermal boundary layer. Direct numerical simulations (DNS) by 
van Reeuwijk et al. (2008a) for Rayleigh numbers up to $10^8$ support systematic deviations from 
a laminar boundary layer on the basis of an analysis of the friction factor and the Reynolds stress budgets. Their DNS showed that the streamwise pressure gradients have a large magnitude compared 
to Reynolds stresses and are not zero as in the Blasius case. Recall also that the active nature of 
the temperature field is not incorporated in the Prandtl-Blasius-Pohlhausen theory. 

Complementary 
to the Prandtl-Blasius-Pohlhausen theory for forced convection similarity solutions for {\em natural convection} are well-known (see e.g. Stewartson 1958; Rotem \& Claassen 1969). Here the buoyancy term remains in the momentum equation (see below) and is balanced by a wall-normal pressure gradient. The temperature differences initiate now fluid motion. Both, purely forced and 
natural convection, were subject to perturbation expansions towards {\em mixed convection} which combines forced and natural convection (Sparrow \& Minkowycz 1962; Leal 1973). 
This means that either the active role of temperature is included as a small-size effect in forced convection or a weak outer flow is imposed in natural convection. Hieber 
(1973) solved numerically the equations which arise from perturbative expansions of forced and natural 
convection. These classical studies are combined with more recent efforts to 
develop two-dimensional boundary layer models for the plume detachment (Fuji 1963; Theerthan 
\& Arakeri 1998; Puthenveetil \& Arakeri 2005; Puthenveetil et al. 2011). The models assume two-dimensional line-like 
thermal plumes  with no significant variation perpendicular to the flow plane.

In this work, we want to resolve the boundary layer structure and its relation to the large-scale 
circulation for $Ra> 10^9$ by means of three-dimensional DNS. 
We aim at better understanding of the physical reasons for the deviations of the 
boundary layer profiles from the classical Prandtl-Blasius-Pohlhausen and Stewartson 
theories for forced and natural convection, respectively. We, therefore, conduct two long-time DNS of turbulent Rayleigh-B\'{e}nard convection in a cylindrical cell at an aspect ratio $\Gamma=1$. 
Step by step it is tested which assumptions of the original derivations of the similarity solutions are satisfied or not. Our studies will include analyses of the LSC, the pressure gradient fluctuations, the importance of violations of the two-dimensionality of the flow and the active role of the 
temperature at the isothermal walls. The coupling between both boundary layers is also analyzed. 
We will show that actually most of the original assumptions that enter all boundary layer theories
are not established in the present cellular flow. Furthermore, we relate locally measured 
turbulence profiles with the results from idealized mixed convection boundary layers.      

The outline of the paper is as follows. In the next section, we summarize the numerical model
and the equations of motion. We present afterwards the boundary layer profiles from the classical time series analysis 
and the dynamical rescaling procedure. The studies are followed by 
investigations of the large-scale circulation, the pressure fluctuations, and time variations of 
the local boundary layer structure. In section 4 we resolve the dynamics in the boundary layer in a small observation window
and relate the findings to results of the boundary layer theory of mixed convection.
We conclude our work with a summary and an outlook.
\begin{figure}
\begin{center}
\includegraphics[width=10cm]{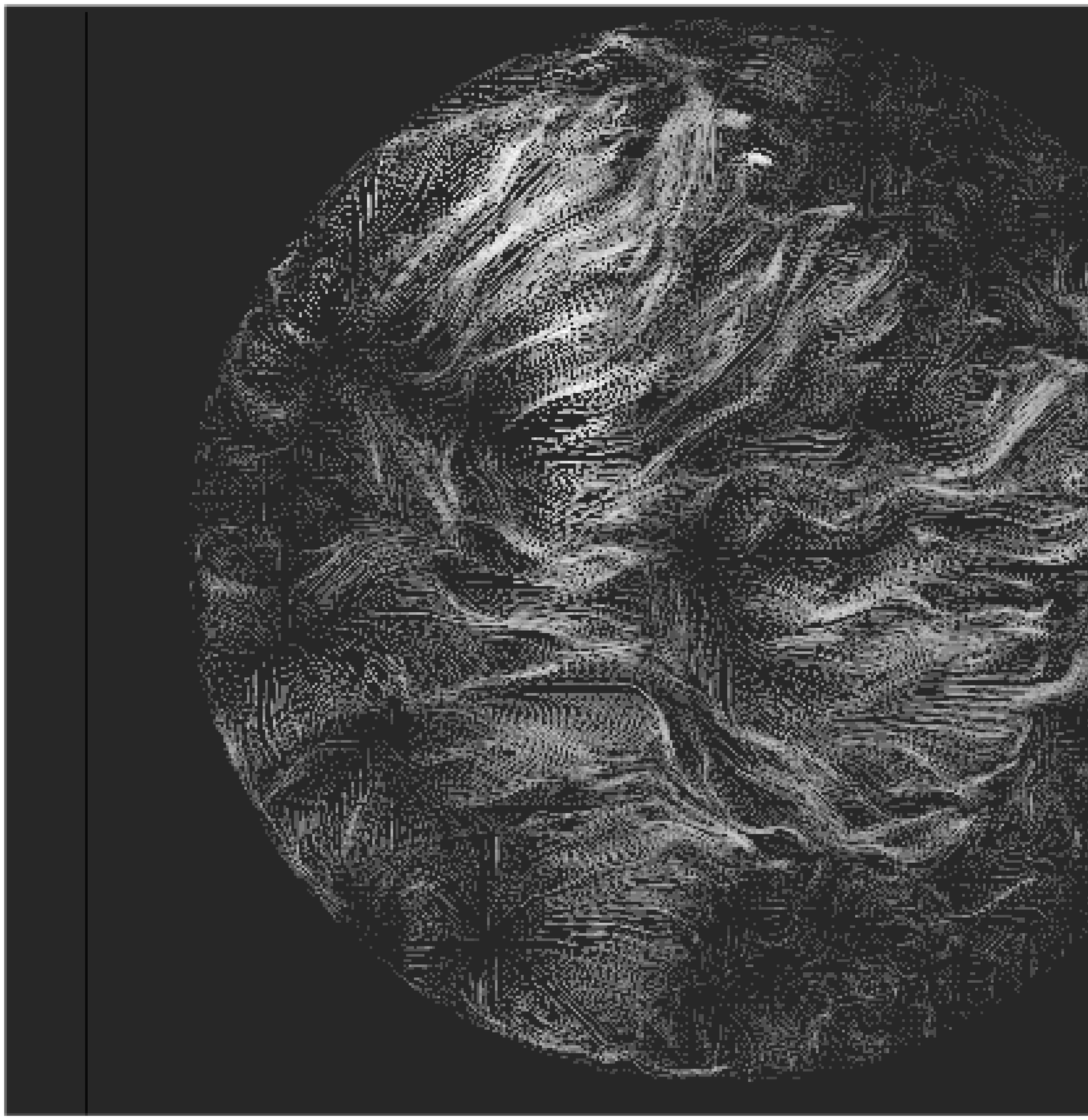}

\vspace{0.5cm}
\includegraphics[width=10cm]{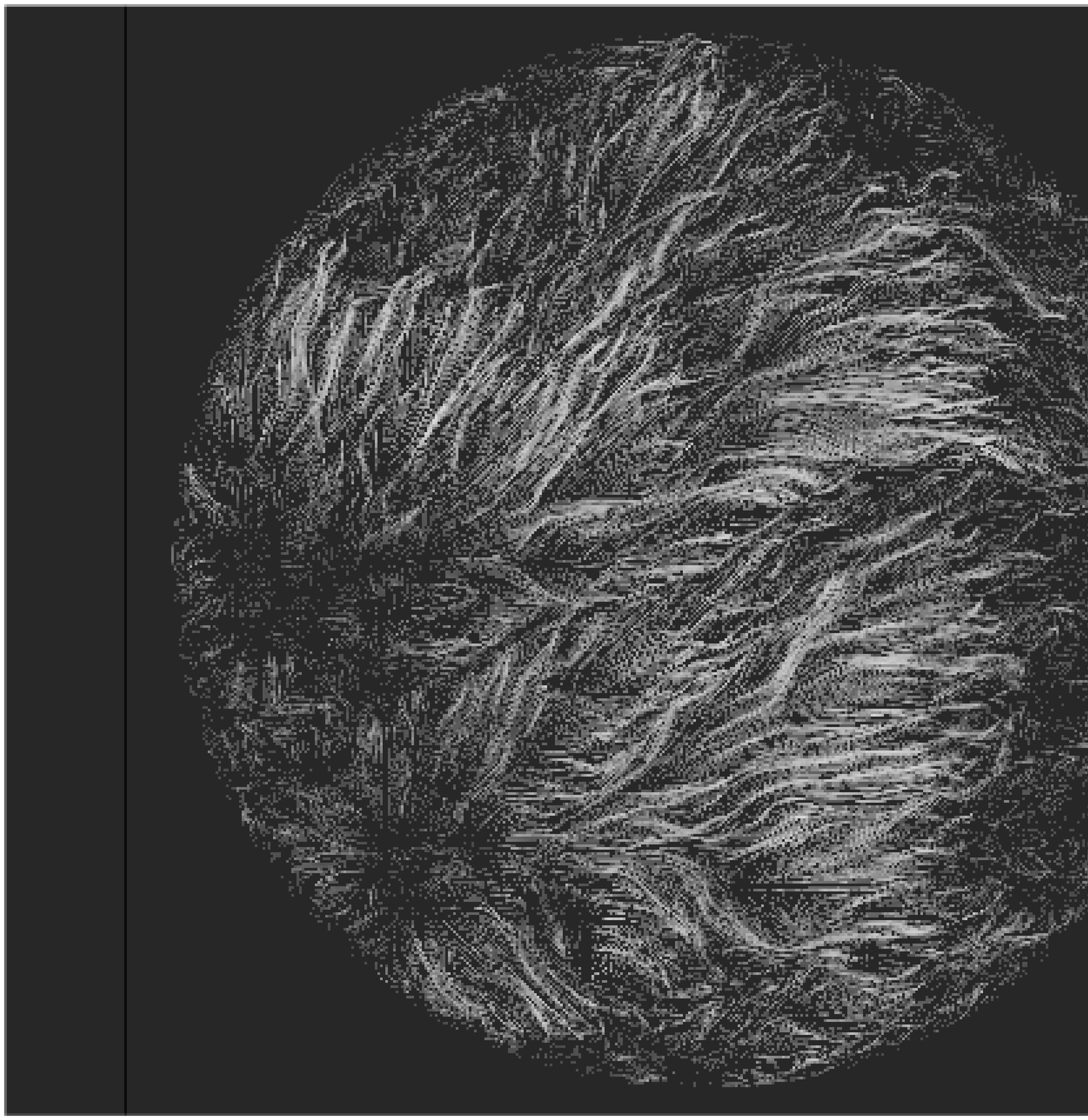}
\caption{Snapshot of three-dimensional stream lines in a turbulent convection cell viewed from the top of boundary layer plane. The lines are seeded in a horizontal plane inside the thermal boundary layer. Top: 
$Ra=3\times 10^9$. Bottom: $Ra=3\times 10^{10}$.} 
\label{fig0}
\end{center}
\end{figure}

\section{Numerical model}
The three-dimensional Navier--Stokes equations in the 
Boussinesq approximation are solved in combination with an advection--diffusion equation
for the temperature field. The system of equations is given by
\begin{eqnarray}
\label{nseq}
\frac{\partial u_i}{\partial t}+u_j \frac{\partial u_i}{\partial x_j}
&=&-\frac{\partial p}{\partial x_i}+\nu \frac{\partial^2 u_i}{\partial x_j^2}+\alpha g T \delta_{iz}\,,\\
\label{ceq}
\frac{\partial u_i}{\partial x_i}&=&0\,,\\
\frac{\partial T}{\partial t}+ u_j\frac{\partial T}{\partial x_j}
&=&\kappa \frac{\partial^2 T}{\partial x_j^2}\,,
\label{pseq}
\end{eqnarray}
where $i,j=x,y,z$. Here $p(x,y,z,t)$ is the kinematic pressure, $u_i(x,y,z,t)$ the velocity field, 
$T(x,y,z,t)$ the total temperature field, $\nu$ the kinematic viscosity, and $\kappa$
the diffusivity of the temperature. The dimensionless control parameters, the Rayleigh number 
$Ra$, the Prandtl number $Pr$, and the aspect ratio $\Gamma$  are defined by
\begin{equation}
Ra=\frac{g\alpha\Delta T H^3}{\nu \kappa}\,,\;\;\;\;\;\;
Pr=\frac{\nu}{\kappa}\,,\;\;\;\;\;\;
\Gamma=\frac{2R}{H}\,.
\end{equation}
Our studies are conducted for $\Gamma=1$, $Pr=0.7$ and  $Ra=3\times 10^9$ and
$3\times 10^{10}$. Constant $\alpha$ is the thermal expansion coefficient, $g$ the gravitational acceleration, $\Delta T$ 
the outer temperature difference, $R$ the radius and $H$ the height of the cylindrical cell. The characteristic length is $H$, 
the characteristic velocity is the free-fall velocity $U_f=\sqrt{g\alpha\Delta T H}$. Times are consequently given in units of 
the free-fall time $T_f=H/U_f$. The cylindrical geometry requires to switch from Cartesian to cylindrical
coordinates, $(x,y,z)\to (r,\phi,z)$. No-slip boundary conditions for the velocity field components , i.e., $u_i\equiv 
0$, hold at all walls. The top and bottom plate are held isothermal at a fixed temperatures
$T_{bottom}$ and $T_{top}$, respectively. The side walls are adiabatic with  $\partial T/\partial r=0$. 
The grid resolutions are $N_r\times N_{\phi}\times  N_z=301\times 513\times 360$ for $Ra=3\times 10^9$ and $513\times 
1153\times 861$ for $Ra=3\times 10^{10}$, where $N_r$, $N_{\phi}$ and $N_z$ are the number of grid points in the radial,  azimuthal and axial directions respectively.

The equations are discretized on a staggered grid with a second-order finite 
difference scheme (Verzicco \& Orlandi 1996; Verzicco \& Camussi 2003). The pressure field $p$ 
is determined by a two-dimensional Poisson solver after applying a one-dimensional Fast Fourier 
Transform (FFT) in the azimuthal direction. The time advancement is done by a third-order 
Runge-Kutta scheme. The grid spacings are non-equidistant in the radial and vertical directions. 
In the vertical direction, the grid spacing is close  to Tschebycheff collocation points. 
The grid resolutions are chosen such that the criterion by Gr\"otzbach (1983) is satisfied plane by 
plane. We define therefore a height-dependent Kolmogorov scale as
\begin{equation}
\eta_K(z)=\frac{\nu^{3/4}}{\langle\epsilon(z)\rangle_{A,t}^{1/4}}\,,
\end{equation}
where the symbol $\langle\cdot\rangle_{A,t}$ denotes an average over a plane at a fixed height $z$ 
and an ensemble of statistically independent snapshots.  Following Emran \& Schumacher (2008) and
Bailon-Cuba et al. (2010),  we define the maximum of the geometric mean of the grid spacing at height $z$ by 
$\tilde{\Delta}(z)=\max_{\Delta_r,\Delta_z}[\sqrt[3]{r\Delta_{\phi}\Delta_r(r)\Delta_z(z)}]$. The 
thermal boundary layer is resolved with 18 grid planes for $Ra=3\times 10^9$ and with 23 grid planes for 
$Ra=3\times 10^{10}$. Thus the recently discussed resolution criterion (Shishkina et al.,  2010), which 
would result in 9 and 13 grid planes for the thermal boundary layer, is satisfied and over-resolved by almost a factor of 2 in both cases. 

The Nusselt number is found to be $Nu=90.32\pm 0.63$ for $Ra=3\times 10^9$ with a standard 
deviation of 0.7\%. The second run at $Ra=3\times 10^{10}$ resulted in $Nu=189.65\pm 1.5$ which gives a 
standard deviation of 0.8\%. The standard deviation is determined in the same way as in Bailon-Cuba et al. (2010).
We take the Nusselt number plane by plane and determine the fluctuation about the global mean.   

Figure \ref{fig0} displays instantaneous 3D velocity fields viewed from the edge of the boundary layer close to the bottom plate for two Rayleigh numbers. Although a preferential mean flow direction is observable, we see significant deviations from a two-dimensionality as visible by the wavy streamlines. With increasing Rayleigh number the streamline plot shows more and more textures on an ever finer scale.   
\begin{figure}
\begin{center}
\includegraphics[width=5cm]{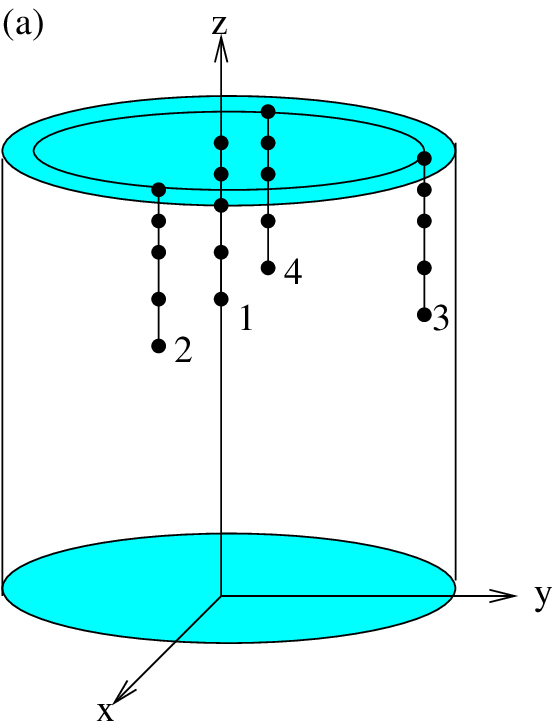}
\hspace{1.6cm}
\includegraphics[width=6.7cm]{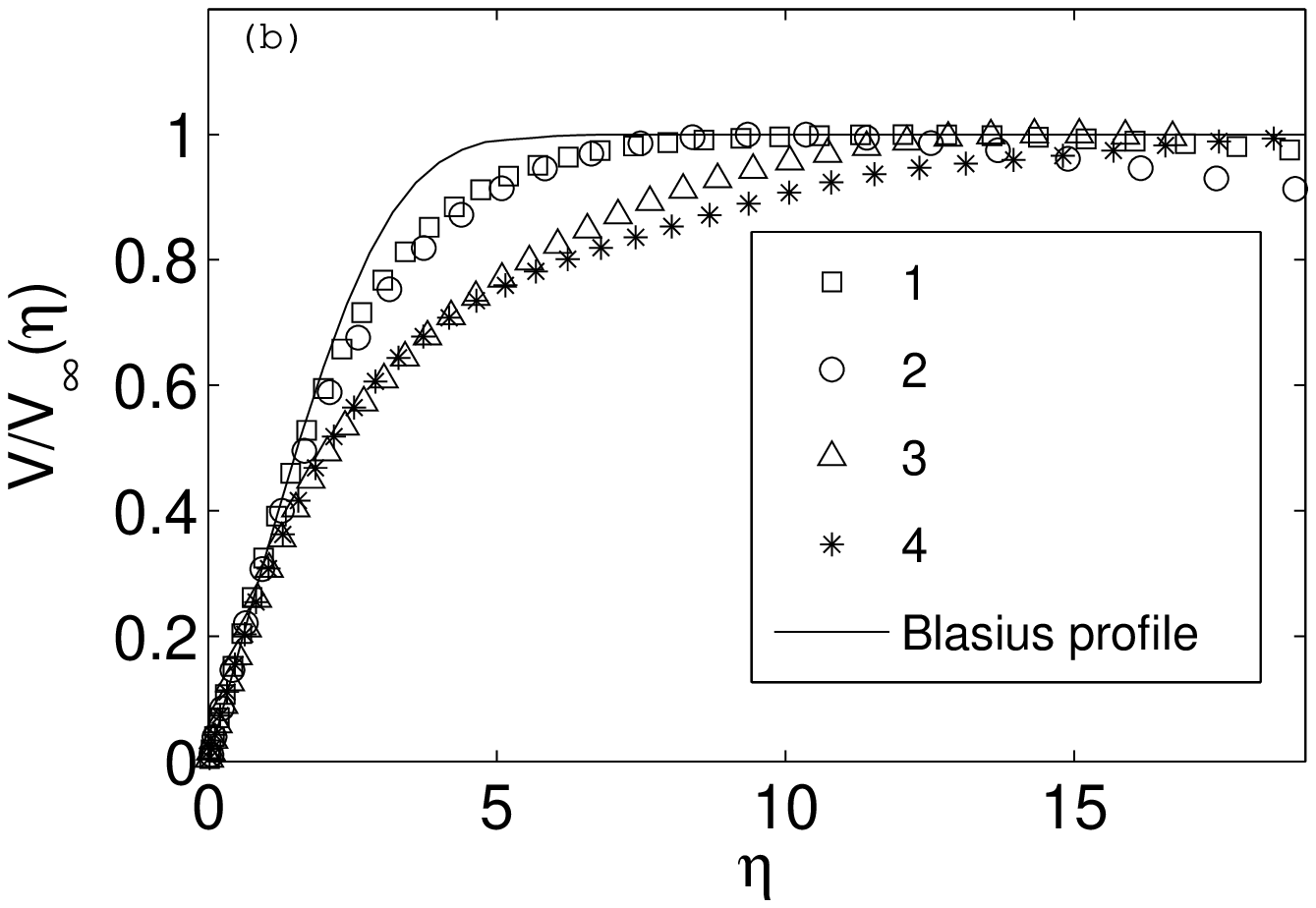}
\includegraphics[width=6.7cm]{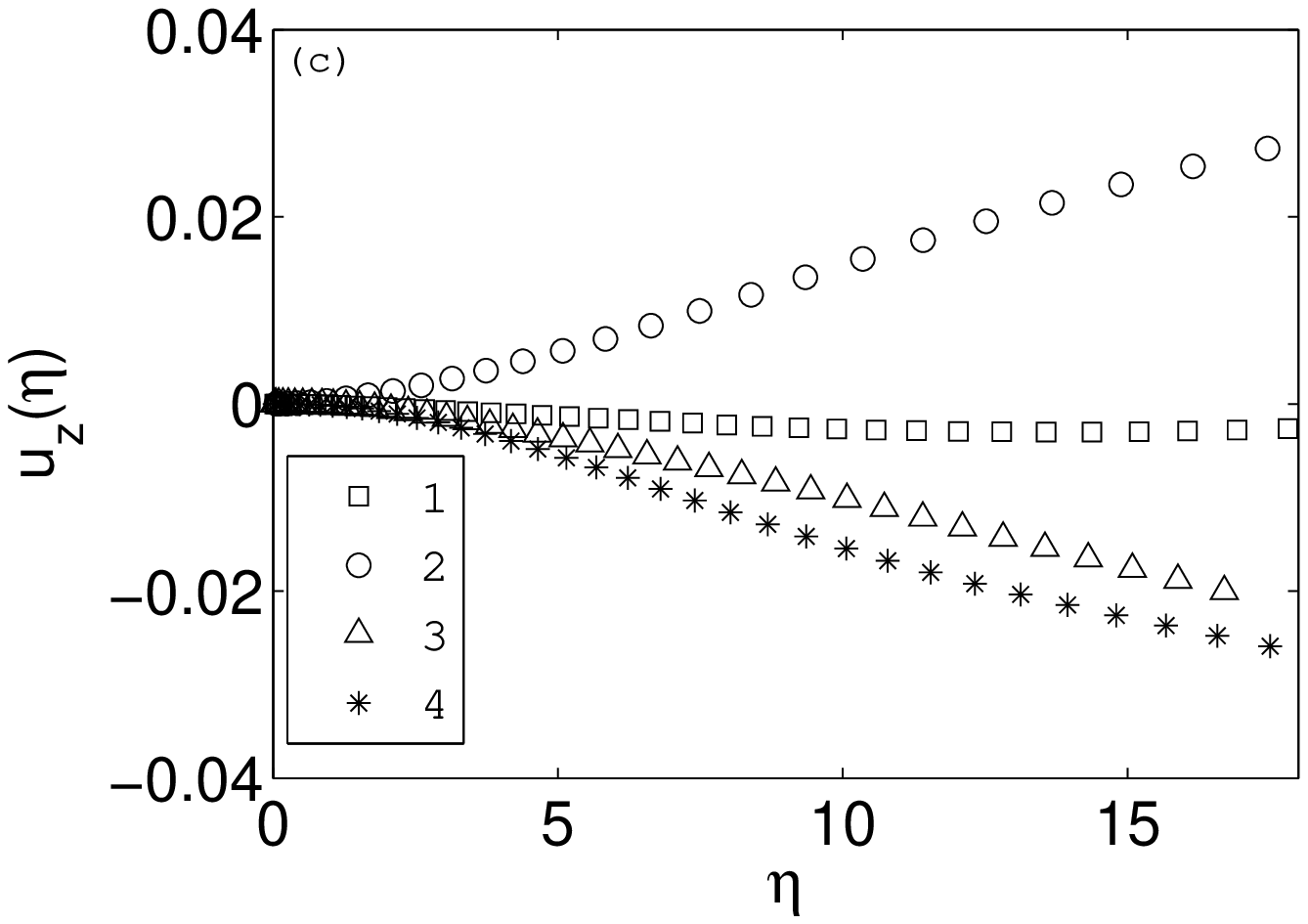}
\includegraphics[width=6.7cm]{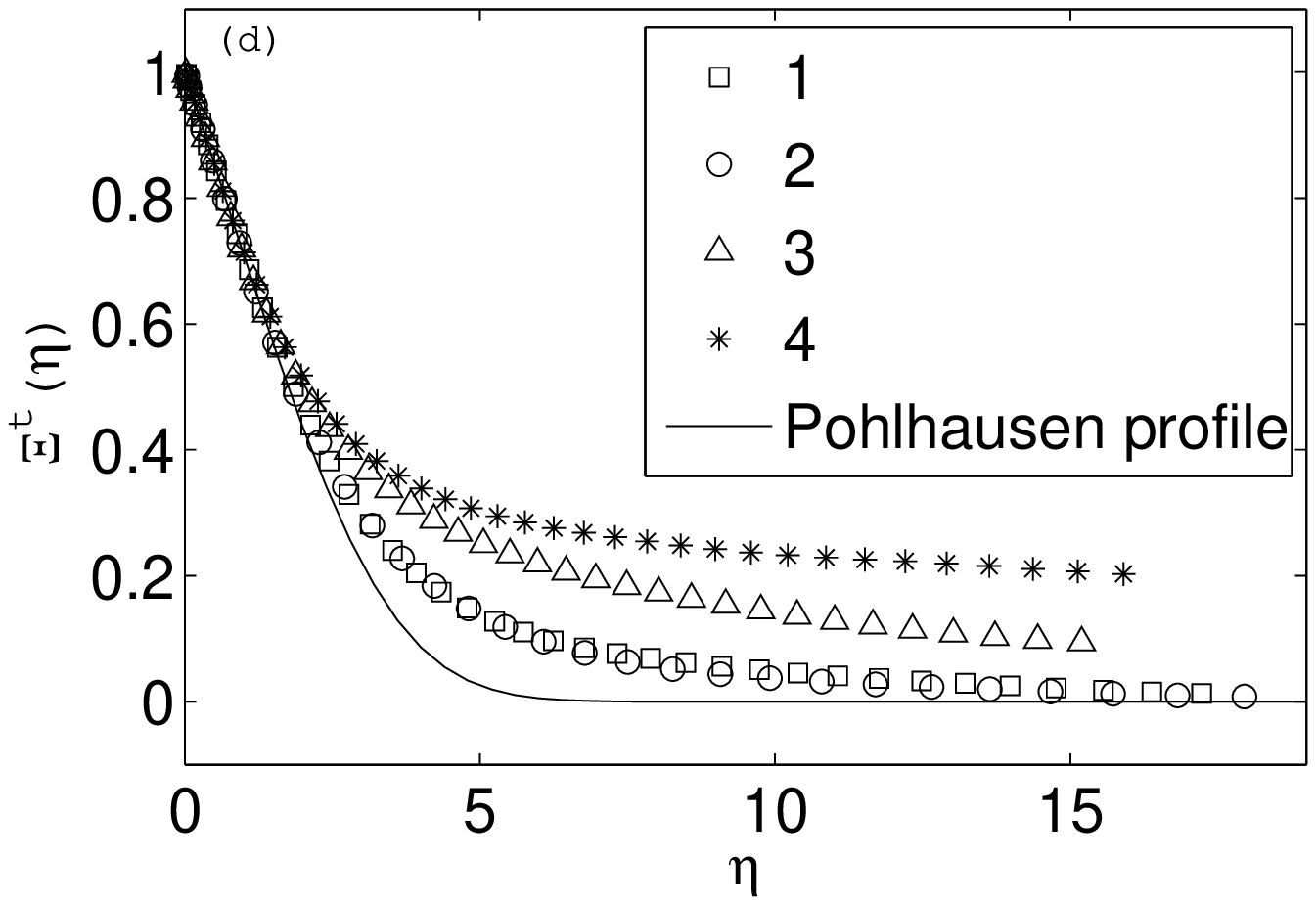}
\caption{(Color online) Mean profiles of velocity and temperature at $Ra=3\times 10^9$. (a) Sketch of the four probe arrays with measurement locations.  
Probe array 1 is mounted at at $(r,\phi)=(0,0)$, array 2 at $(0.88R, 0)$, array 3 at $(0.88R, \pi/2)$ and array 4 at $(0.88 R, \pi)$. (b) Mean profile 
of the horizontal velocity $V(\eta)$ as defined in (\ref{def_v}). (c) Mean profile of the vertical velocity 
component ${u}_z(\eta)$. (d) Mean profile of the rescaled temperature $\Xi^t(\eta)$ which is given by 
(\ref{def_chi}). The solid lines in panels (b) and (d) correspond with the classical Blasius and 
Pohlhausen solutions (Schlichting, 1957).} 
\label{fig1}
\end{center}
\end{figure}
\begin{figure}
\begin{center}
\includegraphics[width=6.7cm]{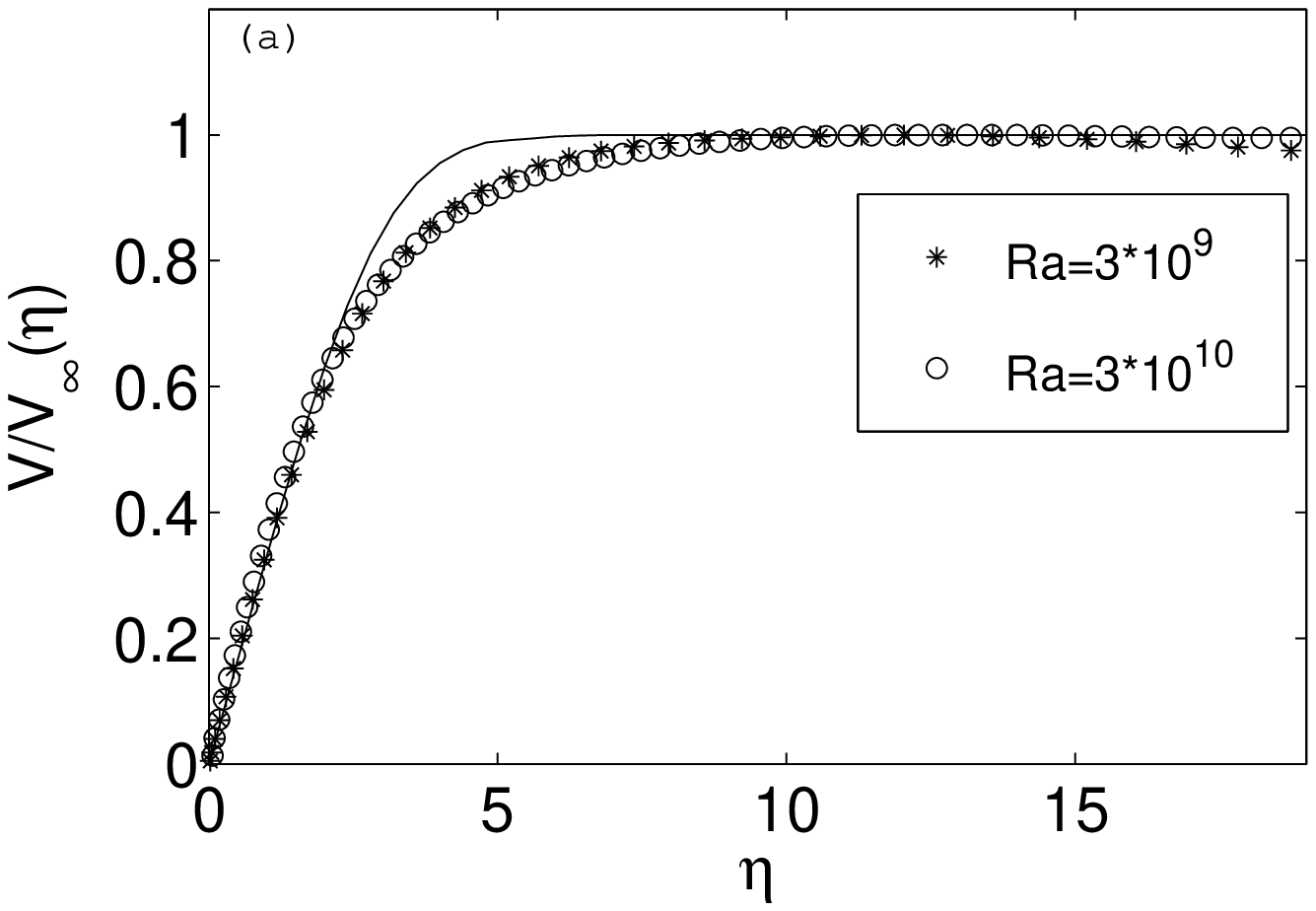}
\includegraphics[width=6.7cm]{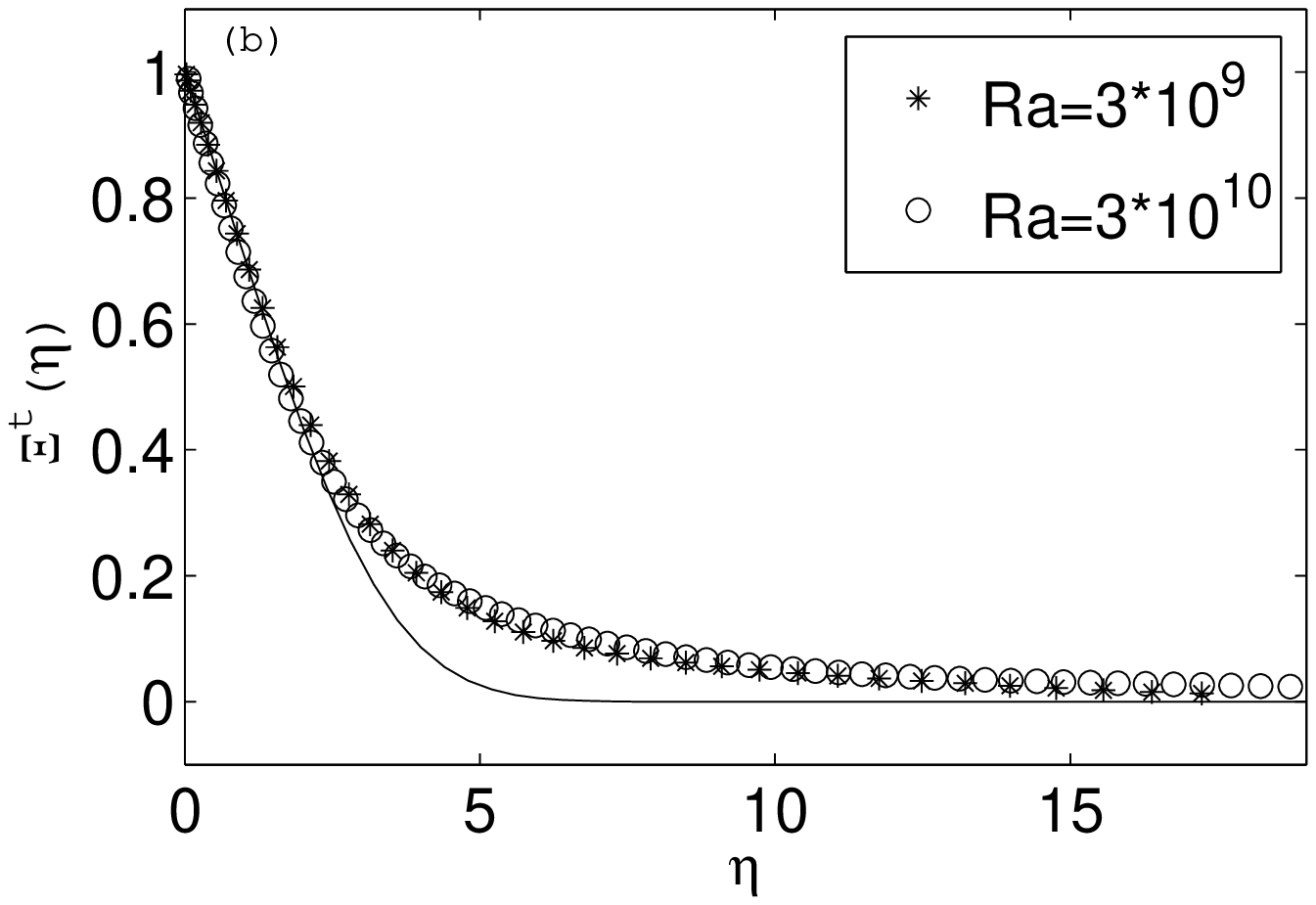}
\caption{Mean profiles of velocity (a) and temperature (b) at probe array 1 for two different Rayleigh numbers. Quantities are the same as in Fig. \ref{fig1}. The solid lines in both panels correspond with the classical Prandtl-Blasius (a) and Pohlhausen solutions (b).} 
\label{Radependence}
\end{center}
\end{figure}
\section{Boundary layer analysis}
\subsection{Vertical mean profiles from time series analysis}
Our numerical approach follows the experimental procedure. The latter consists of measuring time-series of the three velocity components or 
temperature at a given point $(r, \phi, z)$ in the cell,  computing time-averages and repeating the measurement for different values of $z$. 
The result of such procedures are mean profiles of temperature or velocity. In our direct numerical simulation we compute such time series 
simultaneously for an array of  40 (and 100) points starting from $z=H$. Probe array 1 is at the centerline. Probe arrays 2, 3 and 4 are at $r=0.88R$ 
and $\phi=0, \pi/2$ and $\pi$, respectively (see Fig. \ref{fig1}(a)).  We compare the one-dimensional mean profiles for the horizontal velocity $V$ 
(as defined in du Puits et al., 2007) which is given by
\begin{equation}
V(r,\phi,z,t)=\sqrt{u_r^2(r,\phi,z,t)+u_{\phi}^2(r,\phi,z,t)}\,,
\label{def_v}
\end{equation}
the vertical velocity component $u_z$ and the normalized temperatures $\Xi$ from the top (t)  and bottom (b) plates, which are defined as
\begin{eqnarray}
\Xi^{t}(r,\phi,z,t)&=&\frac{T(z=H/2)-T(r,\phi,z,t)}{T(z=H/2)-T_{top}}\,,
\label{def_chitop}\\
\Xi^{b}(r,\phi,z,t)&=&\frac{T(r,\phi,z,t)-T(z=H/2)}{T_{bottom}-T(z=H/2)}\,,
\label{def_chi}
\end{eqnarray}
with the corresponding profiles arising from the Prandtl-Blasius-Pohlhausen theory (see Figs. 
\ref{fig1} (b)-(d)). Here $\eta$ is the similarity variable defined in the appendix in (\ref{Ap_1}). 
The time series contains 57000 data points for $Ra=3\times 10^9$ (and 23000 for $Ra=3\times 10^{10}$) at each position of the 
probe array. This corresponds to 122 (and 58 for $Ra=3\times 10^{10}$) free-fall time units $T_f$. 
Similar to the laboratory experiments by du Puits et al. (2007) and Zhou \& Xia (2010), 
we detect clear deviations from the Blasius and Pohlhausen solutions which are also shown in 
the figures. Furthermore, significant differences among the four profiles can be seen which are 
caused by the existing large-scale flow in the cell. Our profiles at $Ra=3\times 10^9$ suggest that probe array 4 (and probably array 3 as well) 
are significantly altered by a mean downward motion while probe array 2 is the region of mean 
upward motion. The mean downward motion seems to be connected with an increase of the boundary layer thickness as the data relax much slower to the 
Blasius profile. In section 3.4. we will show that the LSC is on average almost perfectly aligned with the $x$-axis ($\phi=0$) for the time interval considered in this particular run. 
In Fig. \ref{Radependence}, we compare the data for the two Rayleigh numbers at the centerline. The 
differences between both data sets are very small. 

\begin{figure}
\begin{center}
\includegraphics[width=6.7cm]{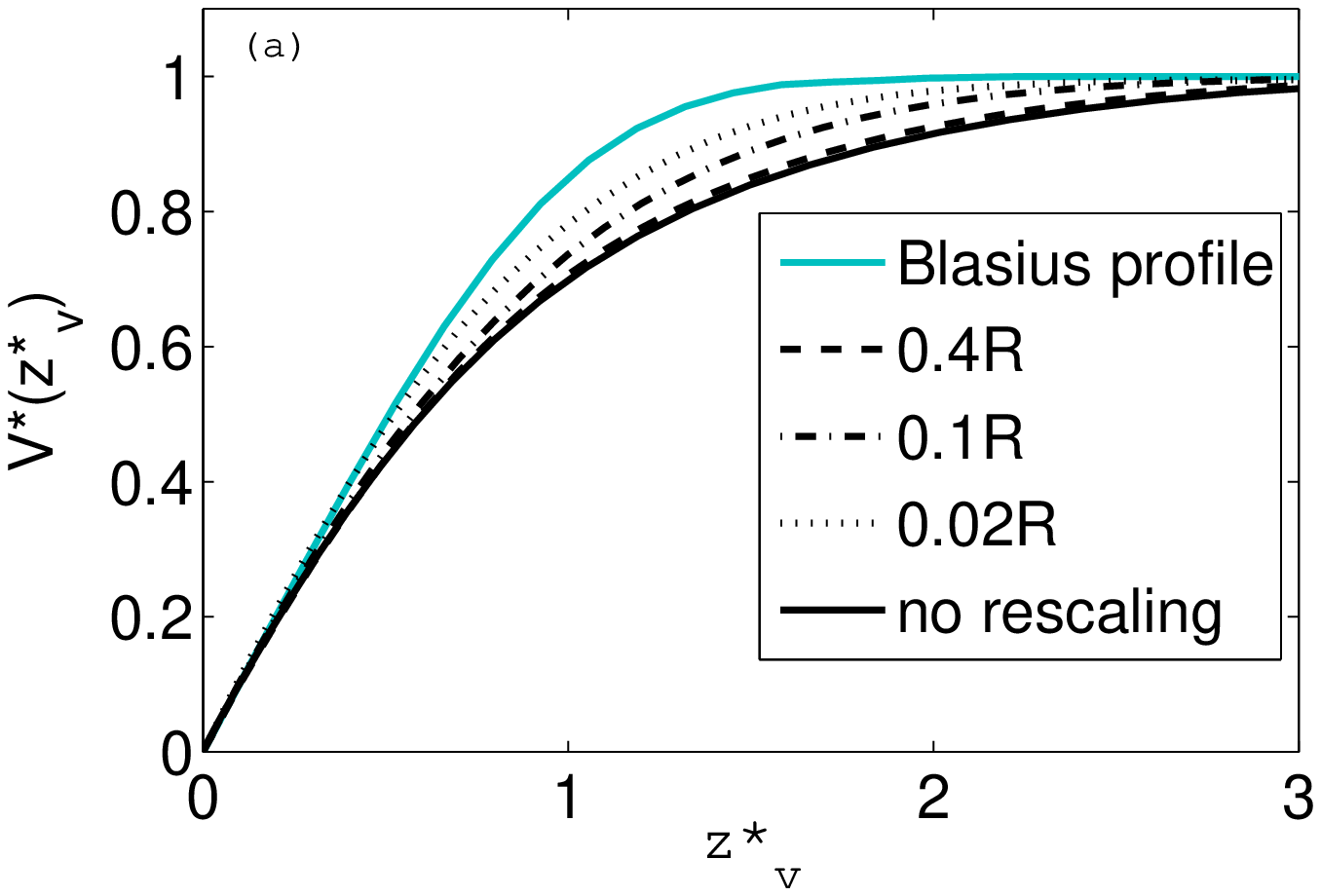}
\includegraphics[width=6.7cm]{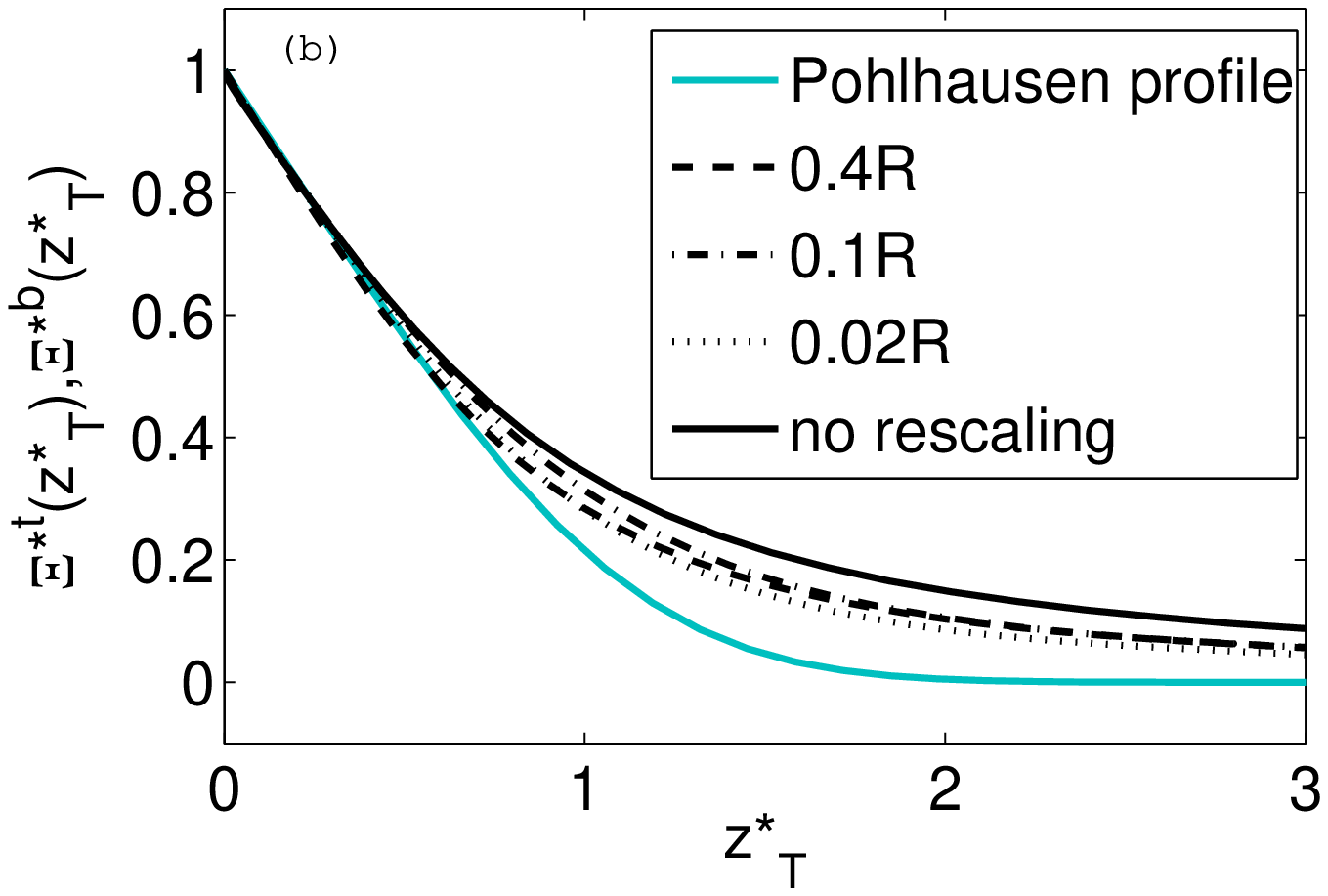}
\includegraphics[width=6.7cm]{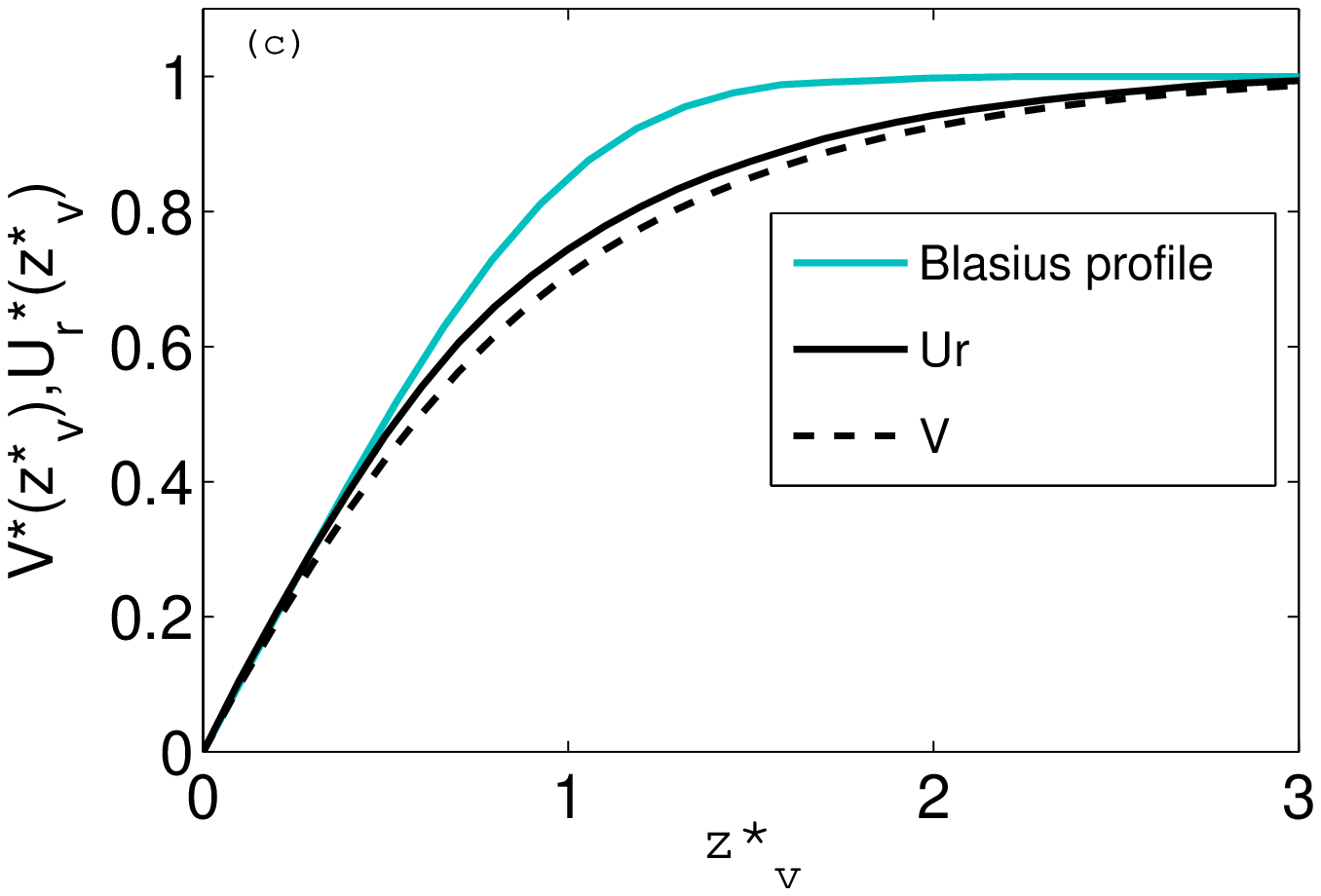}
\includegraphics[width=6.7cm]{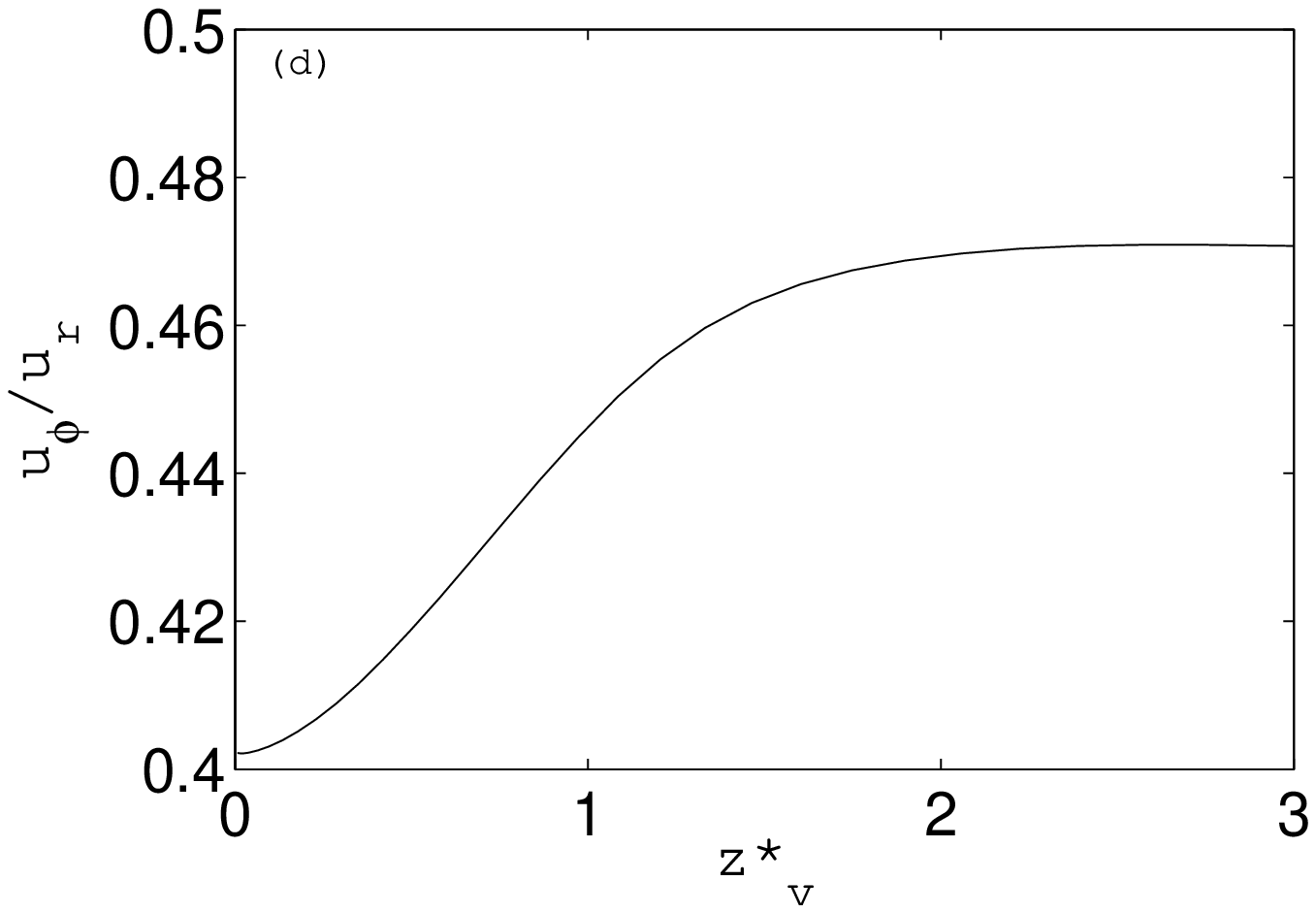}
\caption{(Color online) Dynamic rescaling of the mean profiles. (a) Dynamically rescaled mean 
velocity profile for different widths of the analysis plane. For comparison the Blasius 
profile and the profile without rescaling are added. (b) Dynamically rescaled mean temperature profile. 
Again, the Pohlhausen solution and the mean profile without rescaling are shown additionally. (c) 
Comparison of the rescaling of $V^{\ast}$ as defined in (\ref{Vrescaled}) with
 $U_r^{\ast}(z_v^{\ast})=\langle u_r(r,\phi_{LSC},z,t) | z=z_v^{\ast}\delta_v(t)\rangle_r$ for a window of
 width $0.4R$.  (d) 
Ratio of velocity magnitudes in the plane, $u_r$, and perpendicular to the plane, $u_{\phi}$ for a window of width $0.4R$.}
\label{fig2}
\end{center}
\end{figure}

\subsection{Dynamical rescaling and fluctuations of the boundary layer thickness}
In the next step, we follow the idea of Zhou \& Xia (2010) that was applied in their convection experiment 
in a narrow rectangular cell and investigate if a so-called dynamic rescaling of the boundary layer 
results in mean profiles that come closer to the Prandtl-Blasius-Pohlhausen predictions. 
Similar to the particle image velocimetry in the experiment,  we analyze the fields in a small planar 
window. We take this window in the centre of the cylindrical cell. Zhou et al. (2011) found that the boundary 
layer profiles come closer to the Prandtl-Blasius-Pohlhausen case downstream the LSC. 
This plane is in our case additionally aligned for each snapshot with the direction of the instantaneous 
large-scale wind. This direction is determined by the angle $\phi_{LSC}$ which is 
defined in (\ref{cross_alpha}). The window has a width of $0.02 R$, $0.1 R$ or $0.4 R$ starting from the centerline of 
the cell. In order to improve the statistics, we conduct this analysis at the top and bottom plates 
independently for each snapshot. This implies that the large-scale flow direction has to be 
determined separately at both plates. It is known that the large-scale circulation obeys a slightly 
twisted roll shape (Xi \& Xia 2008). 

The instantaneous velocity boundary layer thickness $\delta_v(t)$ is determined as the 
intersection point of the horizontal line through the first local maximum of the velocity profile 
and the tangent to the profile at the plates. The same procedure is repeated for the 
instantaneous thermal boundary layer thickness $\delta_T(t)$. Vertical distances have to be 
rescaled with 
\begin{equation}
z_v^{\ast}(t)=\frac{z}{\delta_v(t)}\;\;\;\;\;\mbox{and}\;\;\;\;\;z_{T}^{\ast}(t)=\frac{z}{\delta_{T}(t)}\,.
\end{equation}
The resulting velocity or temperature profiles follow by (Zhou \& Xia, 2010)
\begin{eqnarray}
V^{\ast}(z_v^{\ast})&=&\langle V(r,\phi_{LSC},z,t) | z=z_v^{\ast}\delta_v(t)\rangle_r\,,\label{Vrescaled}\\
\Xi^{\ast}(z_T^{\ast})&=&\langle \Xi(r,\phi_{LSC},z,t) | z=z_{T}^{\ast}\delta_{T}(t)\rangle_r\,.
\end{eqnarray}
Here $\langle\cdot\rangle_r$ indicates an averaging with respect to $r$ in the plane that is aligned
in $\phi_{LSC}$ and the rescaled temperature $\Xi$ is taken at the bottom and top, respectively.
The corresponding profiles are shown in Figs. \ref{fig2}(a) to \ref{fig2}(c). Contrary to the experiments 
by Zhou and Xia (2010) and the two-dimensional DNS by Zhou et al. (2010, 2011) deviations to the
Prandtl-Blasius-Pohlhausen profiles remain for all window widths and velocities used. A better  agreement is however observable  when the window is chosen to be narrower in radial extension. A further improvement for the width $0.4R$ is found when the radial component $u_r$ is used  instead of $V$ which is defined in Eq. (\ref{def_v}). For smaller windows, however, the agreement
with respect to $V$ was better again than for $u_r$.
The deviations for the temperature are more persistent which is caused by the plume detachments as we will
see in section 4.  We verified that the results are statistically converged by varying the number of samples. A shift 
of the window away from the center of the plate or a combination of neighboring windows with angles 
around $\phi_{LSC}$ did not lead to a better agreement with the predictions of the Prandtl-Blasius-Pohlhausen 
theory. The same holds for smaller window sizes than $0.02 R$.

A first significant difference to the previous analysis can, however, be identified immediately. 
In Fig. \ref{fig2}(d) we compare the magnitude of the velocity ($u_r$) in the analysis plane 
with the velocity ($u_{\phi}$) perpendicular to the analysis plane with a window of width $0.4R$. It can be seen that the 
ratio takes a significant non-negligible value in contrast to the two-dimensional and quasi-two-dimensional 
situation, respectively.  At $z_v^{\ast}=0.5$, the height for which the measured data 
start to differ from the theoretical profile, the ratio is grown up to  a value of 0.42. This is one important 
difference to the two-dimensional DNS and the quasi-two-dimensional laboratory measurements 
that gives a first hint of why the deviations from the Blasius prediction persist in our geometry.   
\begin{figure}
\begin{center}
\includegraphics[width=13cm]{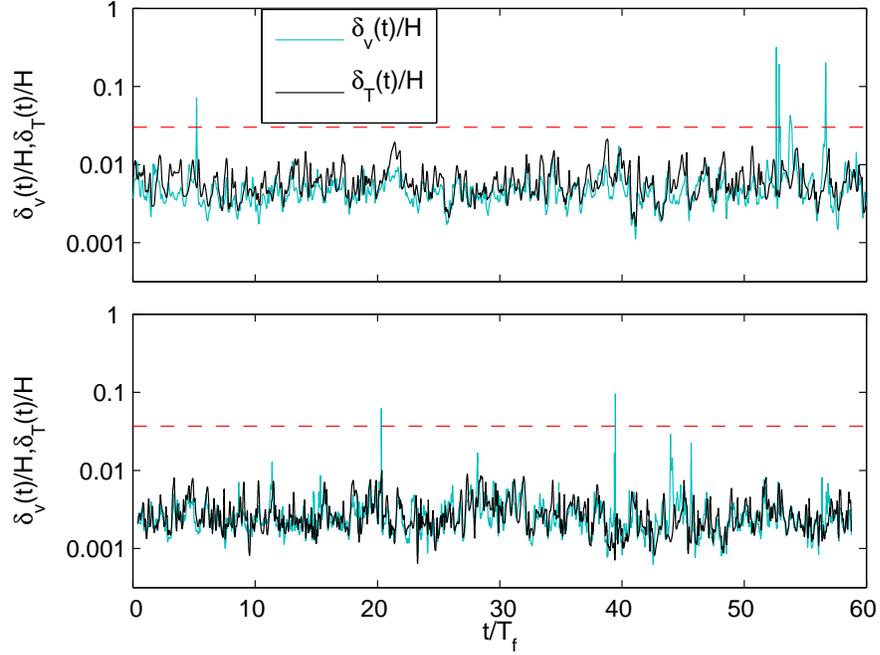}
\caption{(Color online) Fluctuation of the instantaneous thickness of the velocity and thermal 
boundary layers as formed at the center of the top plate of the cell. (top) $Ra=3\times 10^9$. (bottom) $Ra=3\times 10^{10}$. 
The dashed lines in both figures mark the end of probe array 1 at which the data are taken for both $Ra$. Thickness 
values that exceed this height are related with velocity profiles that grow gradually from $z=0$ and therefore an 
intersection point with the horizontal line through the first local maximum that lies beyond the end of the probe array.} 
\label{fig3}
\end{center}
\end{figure}

\subsection{Fluctuating boundary layer thickness} \label{s33}
\begin{figure}
\begin{center}
\includegraphics[width=6.5cm]{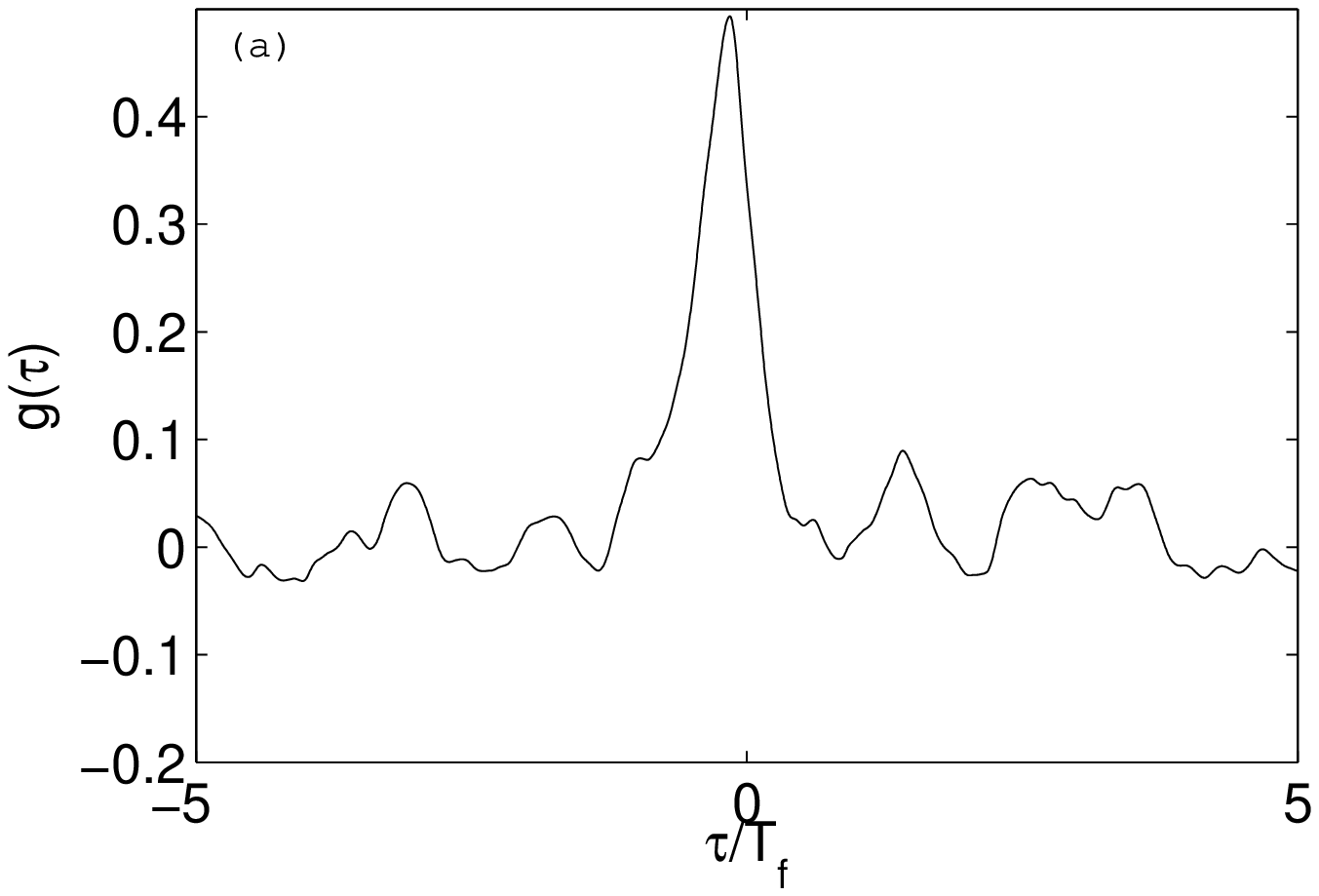}
\includegraphics[width=6.5cm]{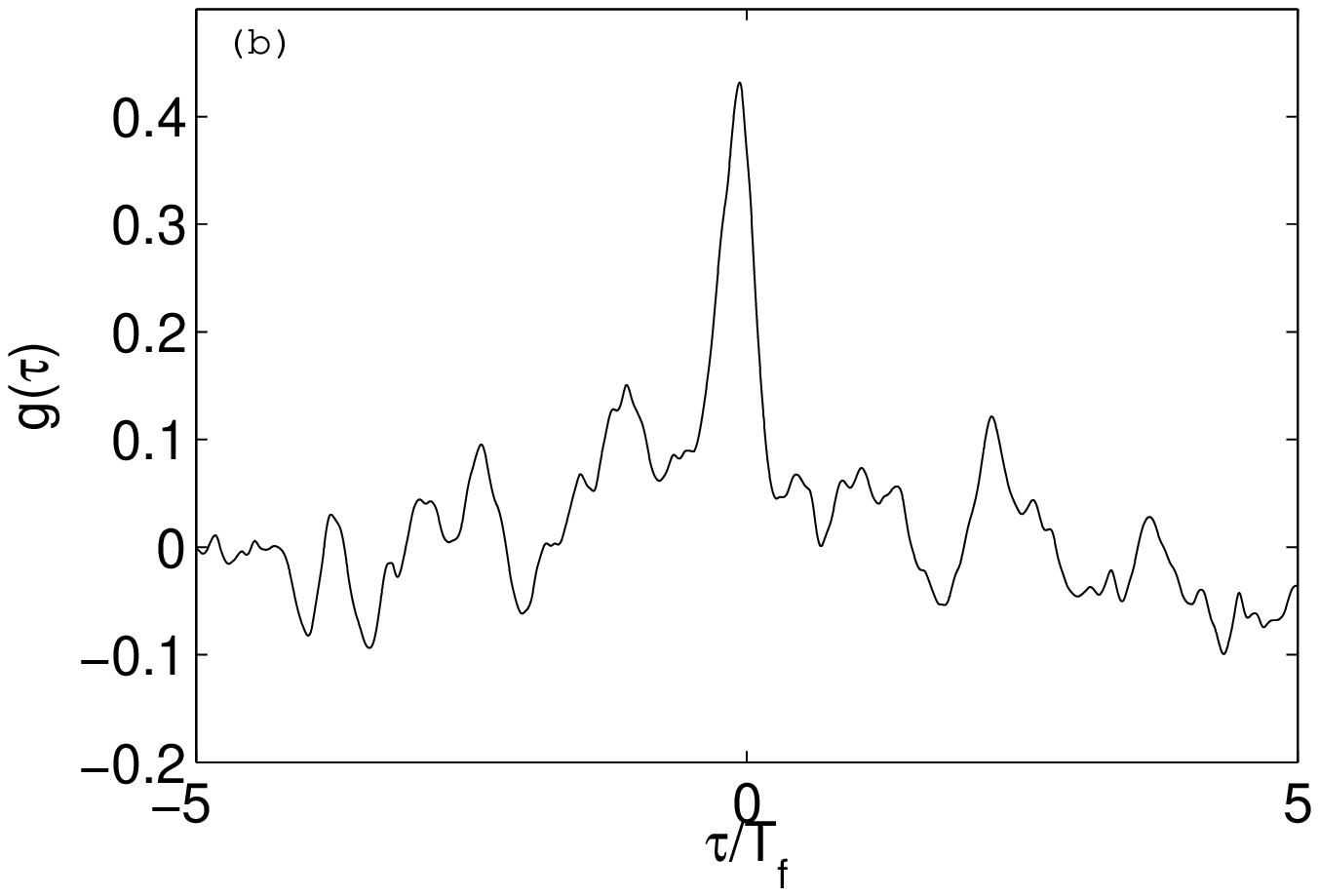}
\caption{Cross-correlation ratio as given in Eq. (\ref{cross}) for the time series of the instantaneous 
boundary layer thickness at the bottom plate. (a) $Ra=3\times 10^9$ with approximately 57000 data points. (b) $Ra=3\times 10^{10}$ with approximately 23000 data points.
The outliers from Fig. \ref{fig3}, i.e. the points that exceed the dashed lines, have been excluded from the analysis.} 
\label{kreuzkorr}
\end{center}
\end{figure}

Figure \ref{fig3} shows time series of both thermal and velocity boundary layer thicknesses obtained from the time 
series at probe array 1. Shorter sequences of same type are obtained from the analysis in the planar observation window which is aligned with the 
instantaneous wind.  We can see that both thicknesses fluctuate strongly for both Rayleigh numbers. Similar to the two-dimensional DNS of Zhou et 
al. (2010), the fluctuations of the velocity boundary layer thickness are slightly stronger than those of the thermal boundary layer. In particular, we
observe rare large thickness events for the velocity which can be related to profiles that grow gradually from $z=0$. We performed a  Fourier 
analysis of both time series and could not detect a characteristic time scale, but  a slowly decaying continuous spectrum which indicates a chaotic 
signal. The cross-correlation ratio which is defined as (Zhou et al., 2010a)
\begin{equation}
g(\tau)=\frac{
\langle [\delta_v(t)-\langle \delta_v\rangle_t]\,[\delta_{T}(t+\tau)-\langle\delta_{T}\rangle_t]\rangle_t
}
{
\sqrt{\langle [\delta_v(t)-\langle\delta_v\rangle_t]^2\rangle_t}\;
\sqrt{\langle [\delta_{T}(t)-\langle\delta_{T}\rangle_t]^2\rangle_t}
}\,.
\label{cross}
\end{equation}
is plotted in Fig. \ref{kreuzkorr} for the fluctuating boundary layer thicknesses at the bottom plate.
The symbol $\langle\cdot\rangle_t$ denotes a time average. Compared to two-dimensional DNS at $Pr=0.7$ (Zhou et al., 2010a), the 
variation of the function $g(\tau)$ is much less regular. In both of our cases the peak is slightly shifted to the left of zero which would indicate 
that variations of the thermal boundary layer cause variations of the velocity boundary layer. The lead time is however shorter as the time that 
we will identify as the time span for a plume detachment. The correlations between both fluctuating 
boundary layers are less pronounced than in the two-dimensional studies. We conclude that such behavior is due to the three-dimensional 
nature of the boundary layer dynamics. We also tried to conduct a similar analysis for the data in the small planes which are aligned with the 
instantaneous large-scale wind. The number of samples was, however, too small for a reliable cross-correlation analysis.  

To extract a characteristic time scale from the time series data, we  analyze the variations of 
boundary layer thicknesses about their means and to determine the average time intervals of 
$\delta_v(t)$ or $\delta_T(t)$ to cross their corresponding means. Our data for $Ra=3\times 10^9$ 
indicate that this interval for both boundary layers is about $T_{cross} \approx 0.5 T_f$ which will turn out to be the time lag 
for a plume detachment in a local region close to the plate. We also repeated the analysis independently for the top boundary 
layers and reproduced this result. A characteristic variation time of the boundary layer is thus $2 T_{cross} \approx T_f$.  
This time can be interpreted as the time at which plumes detach in a local region close to the plates
(see also our analysis in the next section). It is short when compared to the  average loop time of Lagrangian tracers in such a 
cell of approximately 20 $T_f$ (as found in Emran \& Schumacher, 2010) which is a characteristic loop time of the LSC.

\subsection{Fluctuations of the large-scale circulation in the convection cell}
\begin{figure}
\begin{center}
\includegraphics[width=14cm]{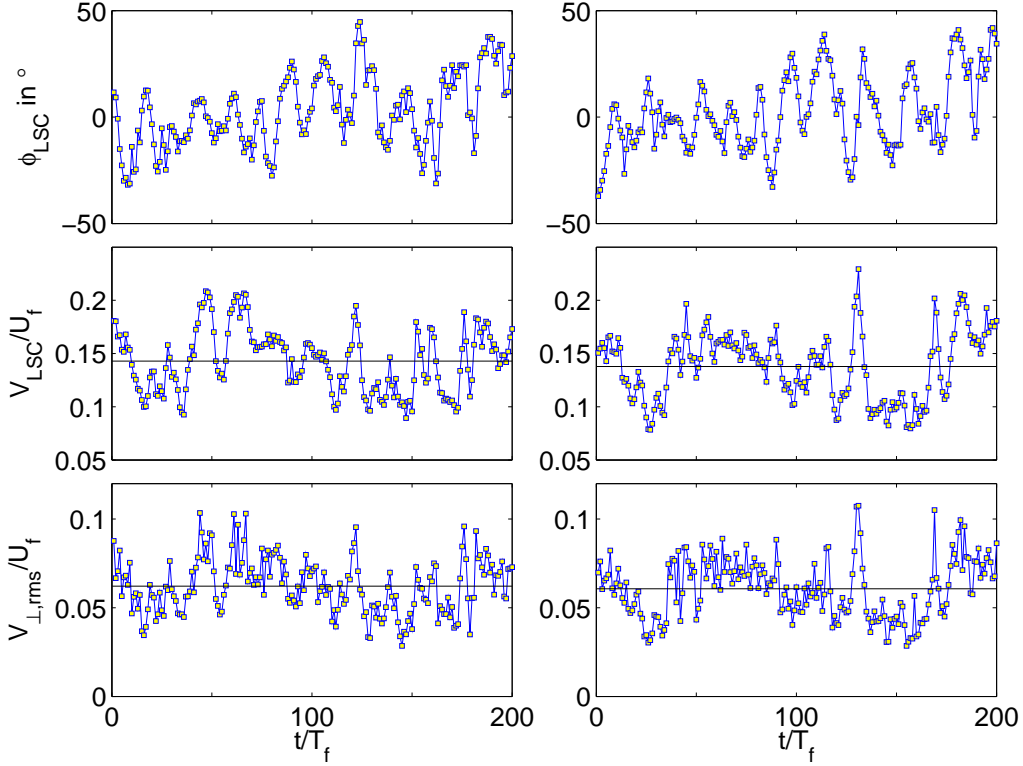}
\caption{(Color online) Direction, magnitude and root mean square of the velocity perpendicular to the instantaneous large-scale circulation. 
Quantities are denoted by $\phi_{LSC}$, $V_{LSC}$ and
$V_{\perp,rms}$. Left column is for the top plate and right column for the bottom plate. Data correspond to the analysis in Fig. \ref{fig2}
and are determined at $z=\delta_T$ for the bottom plate and at $z=H-\delta_T$ for the top plate,
respectively. Data are for $Ra=3\times 10^9$. The solid horizontal lines indicate the means of the time series.}
\label{fig4}
\end{center}
\end{figure}

Figure \ref{fig4} displays the direction (or the angle of orientation) and magnitude of the LSC. The orientation the LSC is used 
for the dynamical rescaling of the boundary layer profiles.  One can see that the orientation of the mean flow at the same instant 
is different at the bottom plate compared to the top plate supporting the idea of a twisted circulation roll (Funfschilling 
\& Ahlers, 2004; Xi \& Xia, 2008; Xi \& Xia, 2008a). The instantaneous direction, $\phi_{LSC}$,  and the magnitude of the large-scale 
circulation, $V_{LSC}$,  are determined by
\begin{eqnarray}
\phi_{LSC}(t_0)&=&\Big\langle\arctan\frac{u_y(x,y,z_0,t_0)}{u_x(x,y,z_0,t_0)}\Big\rangle_{A_r}\,,
\label{cross_alpha}\\
V_{LSC}(t_0)&=&\Big\langle\sqrt{u_x(x,y,z_0,t_0)^2+u_y(x,y,z_0,t_0)^2}\Big\rangle_{A_r}\,,
\label{cross1}
\end{eqnarray}
where the subscript $A_r$ denotes the average over a circular cross-section with $r\le 0.88 R$ 
at $z_0=\delta_T$ for the bottom or $z_0=H-\delta_T$ for the top plate. Furthermore we show the root-mean-square
of the velocity vector which is perpendicular to ${\bf v}=u_x{\bf e}_x+u_y{\bf e}_y$. This crossflow velocity 
vector is determined by the relation ${\bf v}_{\perp}\cdot {\bf v}=0$ at each point $(x,y,z_0)\in A_r$. The quantity is given by 
\begin{equation}
V_{\perp,rms}(t_0)=\sqrt{\langle v_{\perp}^2(x,y,z_0,t_0)\rangle_{A_r}}\,.
\end{equation}
It is seen that the circulation is strongly varying in both amplitude and angle. In case of the angle 
we do observe a fast variation of the orientation over a range of approximately $50^{\circ}$. On average the LSC is
almost perfectly aligned with the $x$-axis ($\phi=0$) along which we have positioned the probe arrays 1,2 and 4. The amount 
of fluctuations perpendicular to the large-scale wind velocity is also significant and reaches up to 50\% of 
$V_{LSC}$. The mean magnitude of $V_{LSC}$ can be used to estimate a LSC turnover time 
by $\tau_{LSC}=\overline{V}_{LSC}^{-1}\times 2\pi(H/2)\approx 21T_f$ which is close to the estimate from
previous Lagrangian studies as mentioned at the end of section 3.3 (Emran \& Schumacher 2010). It is also
consistent with a LSC turnover time of 18 $T_f$ (which corresponds 35 seconds) in the Barrel of Ilmenau.
Furthermore, Ahlers et al. (2009a) report time scale of 25 seconds from their helium experiment at $\Gamma=1/2$
that can be converted into 33 seconds by multiplication with 4/3 for a unit aspect ratio cell. 

On top of  the fast oscillation is a very slow drift of the angle (see panels in the upper row). This indicates  
that a short fraction  of a very slow precession of the large-scale circulation is monitored.
Such a slow-mode can be present since the mean orientation of the roll is not locked in one
particular direction as being frequently  observed in experiments. We are however not able to 
study this slow mode of motion in our DNS since it would exceed our present numerical 
capabilities in terms of the length of the simulation. Better access to this very slow large-scale 
dynamics would require investigations with low-dimensional models (Brown \& Ahlers, 2009) or 
models obtained by proper orthogonal decomposition of the turbulence fields (Bailon-Cuba \& Schumacher, 2011).   
\begin{figure}
\begin{center}
\includegraphics[width=6cm]{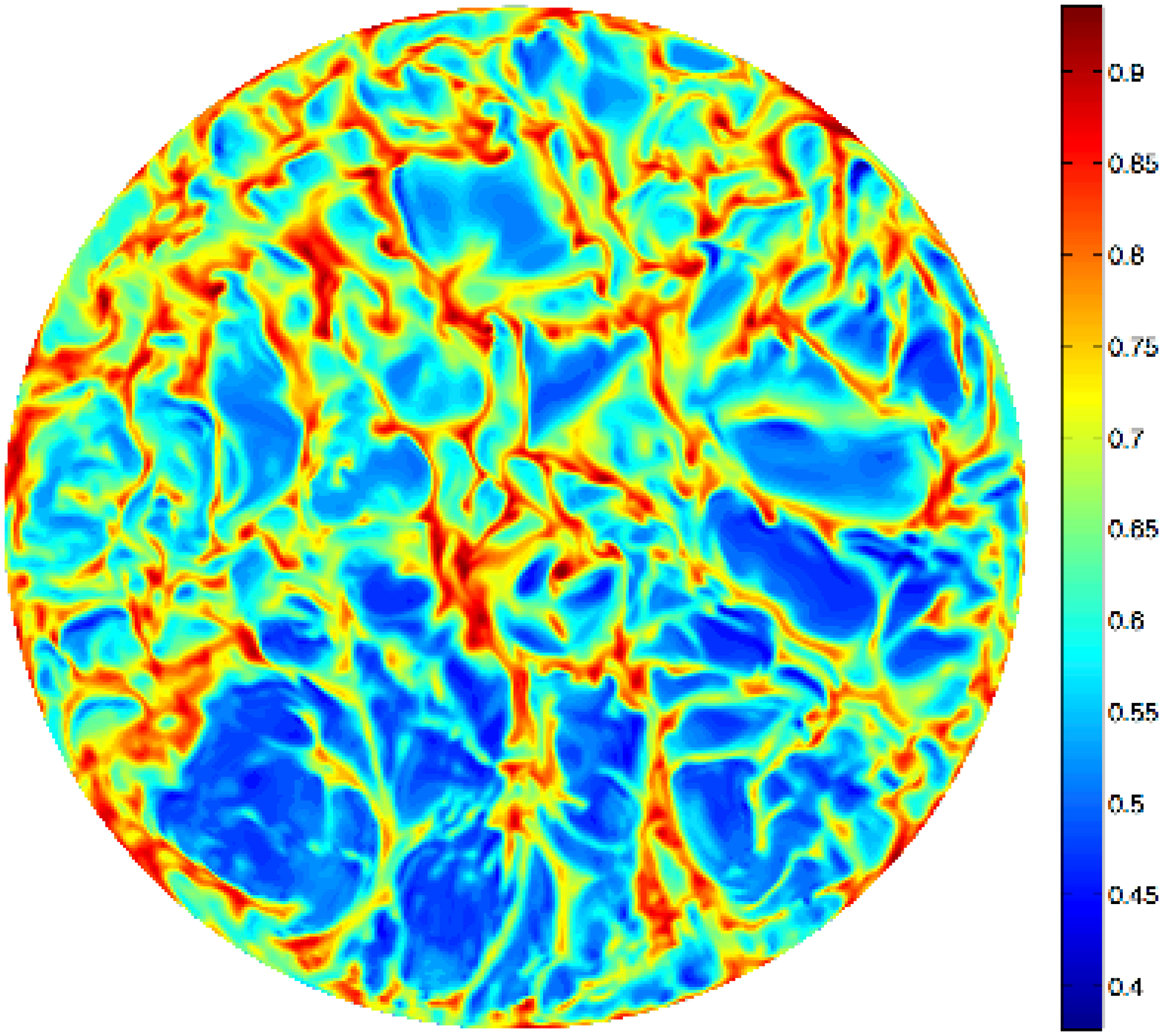}

\includegraphics[width=4.4cm]{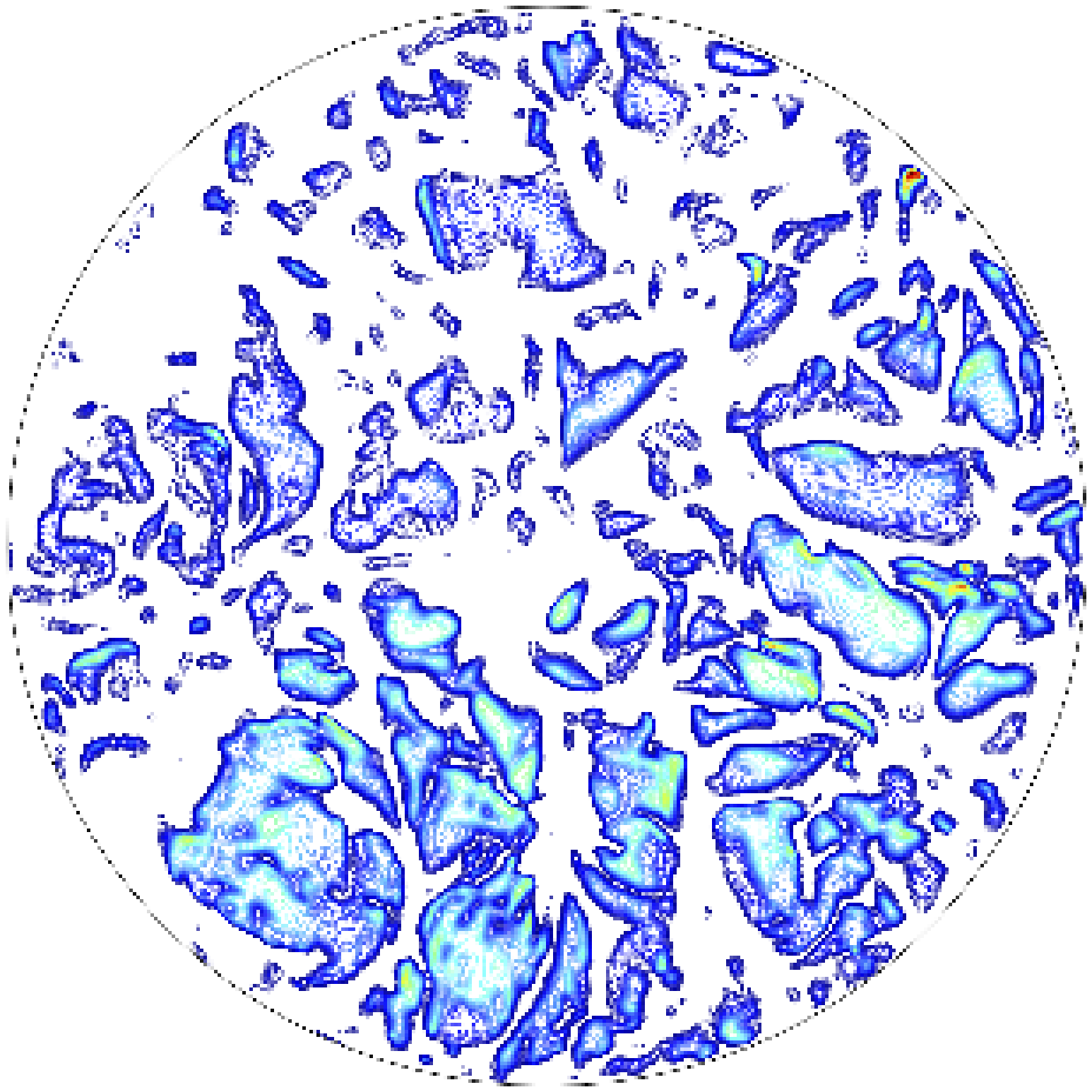}
\includegraphics[width=4.4cm]{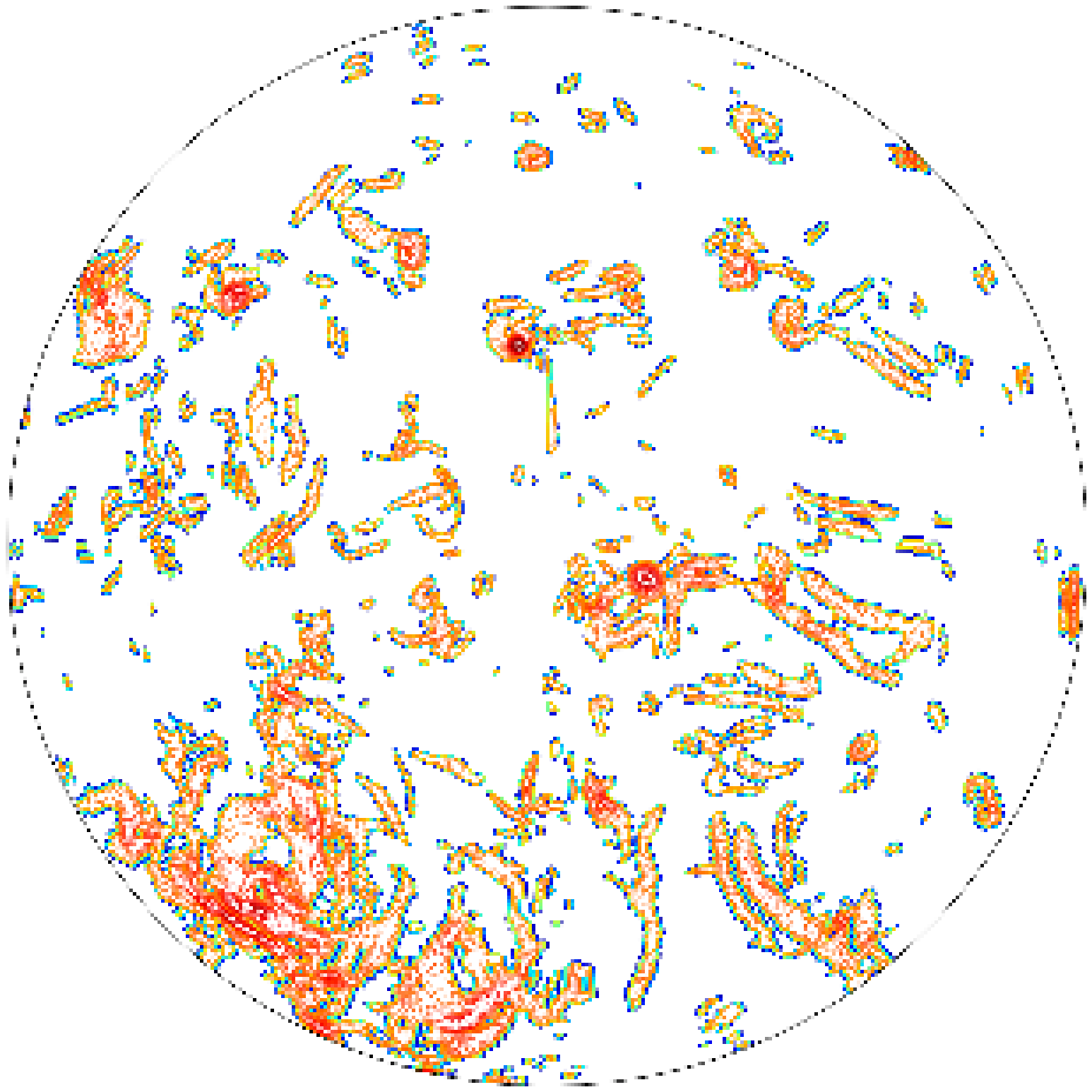}
\includegraphics[width=4.4cm]{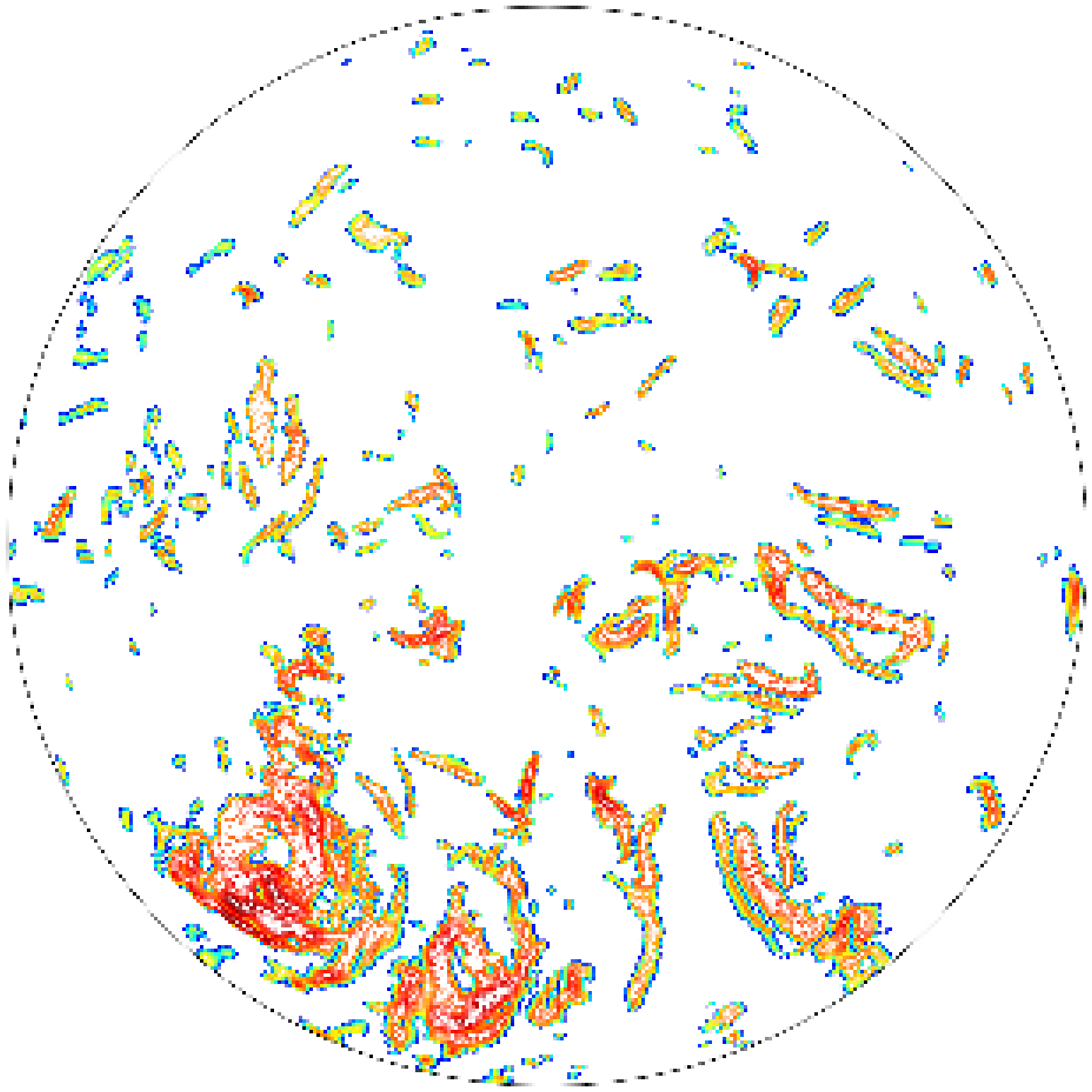}
\caption{(Color online) Spatial correlation between the horizontal pressure gradient and temperature. The top figure shows a horizontal 
cross section of temperature $T$. The three contour plots below show the thresholded temperature fluctuations $T^{\prime}_c$ (left figure), 
the pressure gradient magnitude $\Pi_c$ (mid panel), and the product of both (right panel). Data are for $Ra=3\times 10^9$ and taken
at $z=\delta_T$. Pressure gradient magnitude and product are shown in logarithmic units.}
\label{fig5}
\end{center}
\end{figure}
\begin{figure}
\begin{center}
\includegraphics[width=11cm]{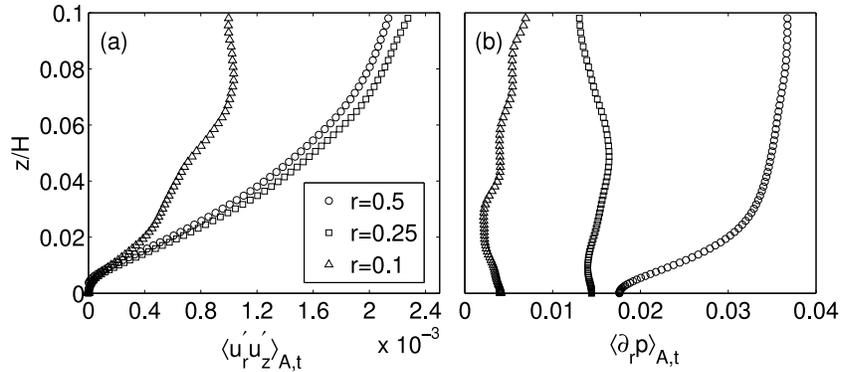}
\caption{Vertical mean profiles of the Reynolds stress (a) and the radial derivative of pressure (b)
obtained over different horizontal cross sections. The full circular cross section $A$ of the cylindrical 
cell with $r=0.5$ (or $r=R$) is compared with smaller cross sections $A$ that have half and one fifth of the radius, all of which concentric with respect to the center line. Data are for $Ra=3\times 10^9$ and in units of $U_f^2$.} 
\label{crosssections}
\end{center}
\end{figure}
\begin{figure}
\begin{center}
\includegraphics[width=13cm]{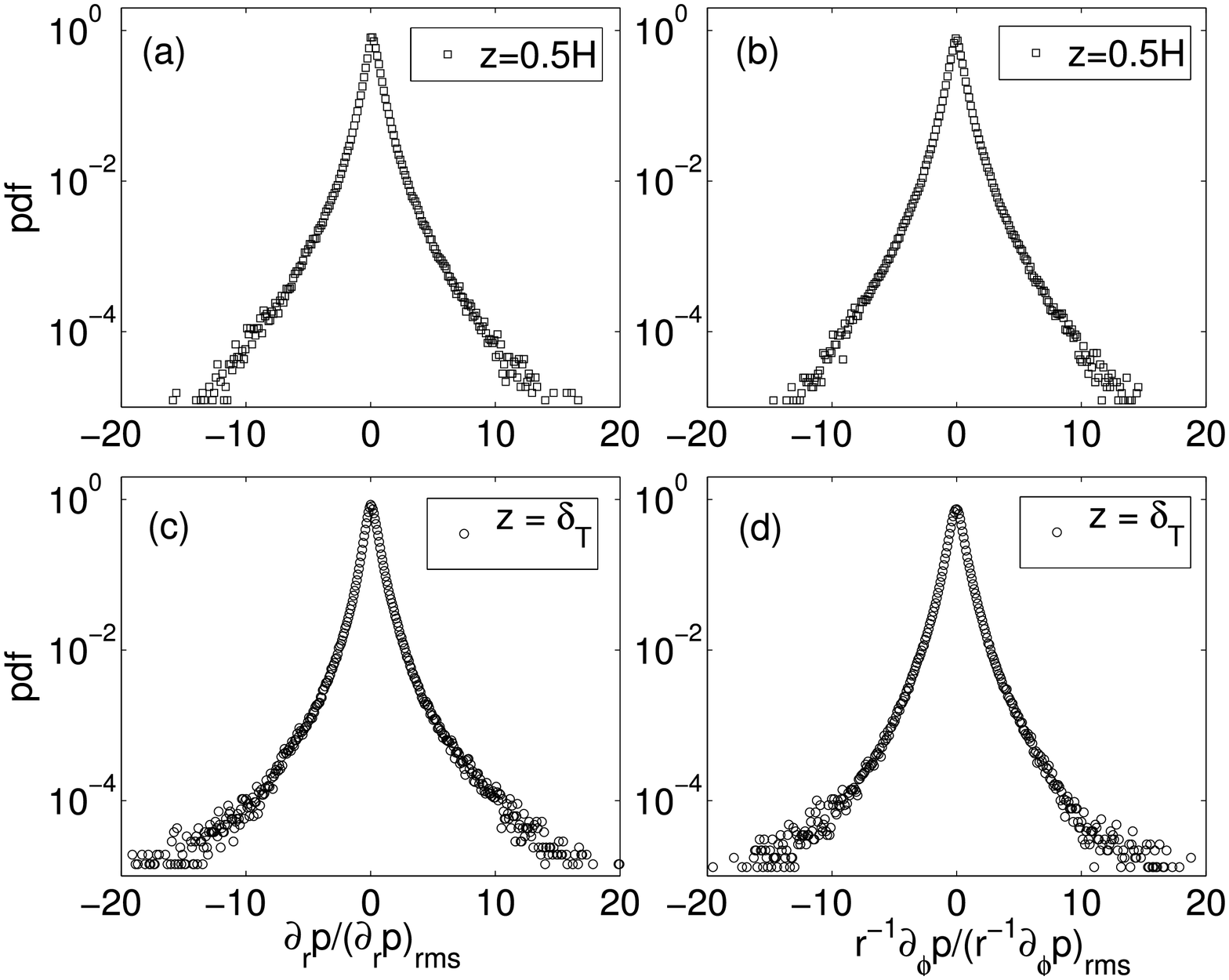}
\caption{Probability density functions of the two horizontal components of the pressure gradient. Figures 
(a) and (c) display $\partial p/\partial r$ in the center plane of the cell and at the top of the boundary layer. Figures (b) and (d) show $r^{-1} \partial p/\partial \phi$. Data are for $Ra=3\times 10^9$.} 
\label{fig6}
\end{center}
\end{figure}

\subsection{Pressure gradient and temperature fluctuations in the boundary layer}
Figure \ref{fig5} shows the temperature, the related temperature fluctuations, which are given by 
\begin{equation}
T^{\prime}({\bf x},t)=T({\bf x},t)-\langle T(z)\rangle_{A,t}\,,
\end{equation}
and the magnitude of the horizontal pressure gradient (mid panel in bottom row),
which is given by
\begin{equation}
\Pi=\sqrt{\left(\frac{\partial p}{\partial r}\right)^2+\left(\frac{1}{r}\frac{\partial p}{\partial\phi}\right)^2}\,.
\end{equation}
Data are taken at the edge of the boundary layer in the plane at $z=\delta_T$ at a time instant 
of the DNS run at $Ra=3\times 10^9$. The contours of $\Pi$, which are plotted in units of the  
logarithm to the base of 10, imply that the pressure field is strongly varying in the horizontal 
plane at this height. More detailed, we display in Fig. \ref{fig5} the quantity 
\begin{equation}
\Pi_c({\bf x},t)=\Pi({\bf x},t)\,\Theta(\Pi-C)\,,
\end{equation} 
with the Heavyside function $\Theta$ and a threshold $C$. The pressure field in the incompressible flow limit is directly 
connected with the flow and thus reflects the high spatial (and temporal) variability  of the flow 
including the large-scale circulation as analyzed in Fig. \ref{fig4}. 

It has been discussed in detail by  Theertan and Arakeri (1998) and Puthenveettil and Arakeri 
(2005) that the horizontal pressure differences are an essential driver of the velocity inside 
the boundary layer. In Fig. \ref{crosssections} we compare vertical profiles taken with respect 
to time and  different horizontal cross sections $A=2\pi r$. Averages of the  radial component of the pressure gradient 
$\langle \partial p/\partial r\rangle_{A,t}$ and the Reynolds stress $\langle u_r^{\prime} u_z^{\prime}\rangle_{A,t}$ 
are shown as examples. The pressure gradient component is non-negligible in the boundary layer. 
Similar to van Reeuwijk et al. (2008a), we compare here the ratio 
\begin{eqnarray}
\gamma=\frac{\Big|\int_0^{\delta_v} \langle\partial p/\partial r (z)\rangle_{A,t}\,\mbox{d}z\Big|}
{\Big|\langle u_r^{\prime}u_z^{\prime}\rangle_{A,t}\Big|_{\delta_v}}\,.
\label{ratio}
\end{eqnarray}
Note that both terms contribute to the friction factor. Values of $\gamma=1.16,\,1.77$ and 5.21 were 
obtained for cross sections $A$ with radius $R$, $R/2$ and $R/5$. We thus confirm 
their finding that this ratio is significant even in a central region where the data come closest to the 
Blasius profiles. We recall that the pressure gradient would be zero in the Blasisus case. 

When the spatial support of $\Pi_c$ is compared with the temperature distribution in the same 
horizontal plane (see bottom row of Fig. \ref{fig5}) we observe that maxima of $\Pi$ are found mostly in the low-temperature voids in between 
the skeleton of plumes, i.e. in regions which are given by (see left panel in bottom row) by $T^{\prime}<0$ or 
\begin{equation}
T^{\prime}_c({\bf x},t)=T^{\prime}({\bf x},t)\,\Theta(-T^{\prime})\,. 
\end{equation}
Again, we use the Heavyside function $\Theta$. In regions of high pressure gradient the 
horizontal flow will be accelerated and piles up local plumes that eventually detach from the boundary layer. The spatial correlation becomes
directly visible when both thresholded fields $\Pi_c$ and $T^{\prime}_c$ are multiplied as shown in the right panel of the bottom row of Fig. \ref{fig5}.
The area covered by these correlated regions is about 11 per cent of the total area and remained nearly constant in time, which we verified by a
pressure field snapshot analysis over a few free fall time units in case of $Ra=3\times 10^9$.

It is also observed from the top panel of Fig. \ref{fig5}, that the plumes are line-like, however with significant thickness modulations along their stems. At Prandtl number 0.7 and for the present Rayleigh numbers, diffusion still plays an important role in the plume formation. This will to our view also result in a limited applicability 
of two-dimensional plume models in which spatial variations in the third direction are assumed to be small (e.g. Puthenveetil \& Arakeri 2005). 
Similar temperature patterns have been found in Zhou et al. (2007), Shishkina \& Wagner (2008), Zhou \& Xia (2010b) and Puthenveetil et al. (2011) where length, width and aspect ratio of the filaments in this skeleton of plumes have been quantified in detail. 

Figure \ref{fig6} displays the probability density function (PDF) of the two components of the pressure gradient in two planes parallel to the bottom plate. This figure underlines the findings from Fig. \ref{fig5}. The fluctuations of the pressure gradient across the boundary layer are highly intermittent as shown by the stretched exponential probability density functions of the both components. In Emran \& Schumacher (2008), the statistics of the temperature field and its gradients has been studied in detail. The spatial variations of the temperature as quantified by the statistics of the temperature gradient components as well as the thermal dissipation rate were found to obey the strongest spatial intermittency in the boundary layer. The intermittency of the pressure gradient field shows 
qualitatively the same behavior, it is enhanced in the boundary  layer.    

We summarize our boundary analysis at this point. The numerical data demonstrate that significant differences from the classical Prandtl-Blasius-Pohlhausen theory are pre\-sent in comparison to the 
two-dimensional case and the quasi-two-dimensional experiments. The near-wall flow and 
temperature structures  are three-dimensional and unsteady as the large-scale circulation to which
the boundary layer dynamics is coupled. This is in line with a fluctuating large-scale circulation and the horizontal pressure gradient in the cylindrical cell.  

\section{Comparison with laminar boundary layers of mixed convection}
\subsection{Two-dimensional boundary layer theory of mixed convection}
As already discussed in the introduction, the boundary layer in turbulent convection can be 
considered as mixed type, i.e. driven by the natural convection and additionally by the LSC. In the classical  boundary layer theory both limiting cases have been studied to some extension. These are the purely forced convective flow also known as the classical Prandtl-Blasius-Pohlhausen case (Blasius 1908; Pohlhausen 1921) and the purely natural convective flow (Stewartson 1958; Rotem \& Claassen 1969). 
For mixed convection, the  Boussinesq equations of motion (\ref{nseq})--(\ref{pseq}) are reduced to the following 
set of two-dimensional and steady boundary layer equations (Schlichting 1957)
\begin{eqnarray}
\label{bleq1}
u_x \frac{\partial u_x}{\partial x}+u_z \frac{\partial u_x}{\partial z}
&=&-\frac{\partial p}{\partial x}+\nu \frac{\partial^2 u_x}{\partial z^2}\,,\\
\label{bleq2}
0 &=&-\frac{\partial p}{\partial z}+\alpha g T \,,\\
\label{bleq3}
\frac{\partial u_x}{\partial x}+\frac{\partial u_z}{\partial z}&=&0\,,\\
\label{bleq4}
u_x\frac{\partial T}{\partial x}+u_z\frac{\partial T}{\partial z}
&=&\kappa \frac{\partial^2 T}{\partial z^2}\,,
\end{eqnarray}
The corresponding dimensionless parameters are the Reynolds and Grashof numbers of the problem which are given by
\begin{equation}
\label{Re1}
Re_x=\frac{V_{\infty} x}{\nu}\,,\;\;\;\;\;\;\;\;Gr_x=\frac{g\alpha (T_w-T_{\infty}) x^3}{\nu^2}\,.
\end{equation}
At the plate ($z=0$) the boundary conditions are $T=T_w$ and $u_x=u_z=0$. Far away from the plate ($z\to\infty$) it follows that $T=T_{\infty}$ and 
\begin{equation}
u_x(z\to\infty)= \left\{
\begin{array}{ll}
V_{\infty} & 
             \;\;\;\text{for forced convection}\\
             0    &  
             \;\;\;\text{for natural convection}
\end{array} \right.
\label{bc1}
\end{equation}
In both cases one can define similarity variables $\eta$ and parameters $\epsilon$ for the perturbation expansion of mixed convection.
In agreement with the definitions (\ref{def_v}) -- (\ref{def_chi}) we can proceed as follows. 
Starting from purely forced convection, the expansion follows to (Sparrow \& Minkowycz, 1962)
\begin{eqnarray}
\frac{u_r(x,z)}{V_{\infty}}&=&f_0^{\prime}(\eta)+\epsilon f_1^{\prime}(\eta)+\dots\,,\\
\Xi(x,z)&=&\frac{T(x,z)-T_{\infty}}{T_w-T_{\infty}}=\theta_0(\eta)+\epsilon \theta_1(\eta)+\dots \,,
\end{eqnarray}
while starting from purely natural convection, it reads (Stewartson, 1958)
\begin{eqnarray}
\frac{u_r(x,z)}{V_n(x)}&=&g_0^{\prime}(\eta)+\epsilon g_1^{\prime}(\eta)+\dots\,,\\
\Xi(x,z)&=&\frac{T(x,z)-T_{\infty}}{T_w-T_{\infty}}=\chi_0(\eta)+\epsilon \chi_1(\eta)+\dots \,,
\end{eqnarray}
where functions with index 0 represent the unperturbed velocity components or temperature. Furthermore $V_{n}(x)=
(\nu g^2 \alpha^2 (T_w-T_{\infty})^2 x)^{1/5}$. More details are provided in the 
appendix for completeness. The resulting systems of perturbation equations for the boundary value problems can be solved 
by a shooting method using a 4th-order Runge-Kutta scheme (Hieber 1973). 

\begin{figure}
\begin{center}
\includegraphics[width=6.7cm]{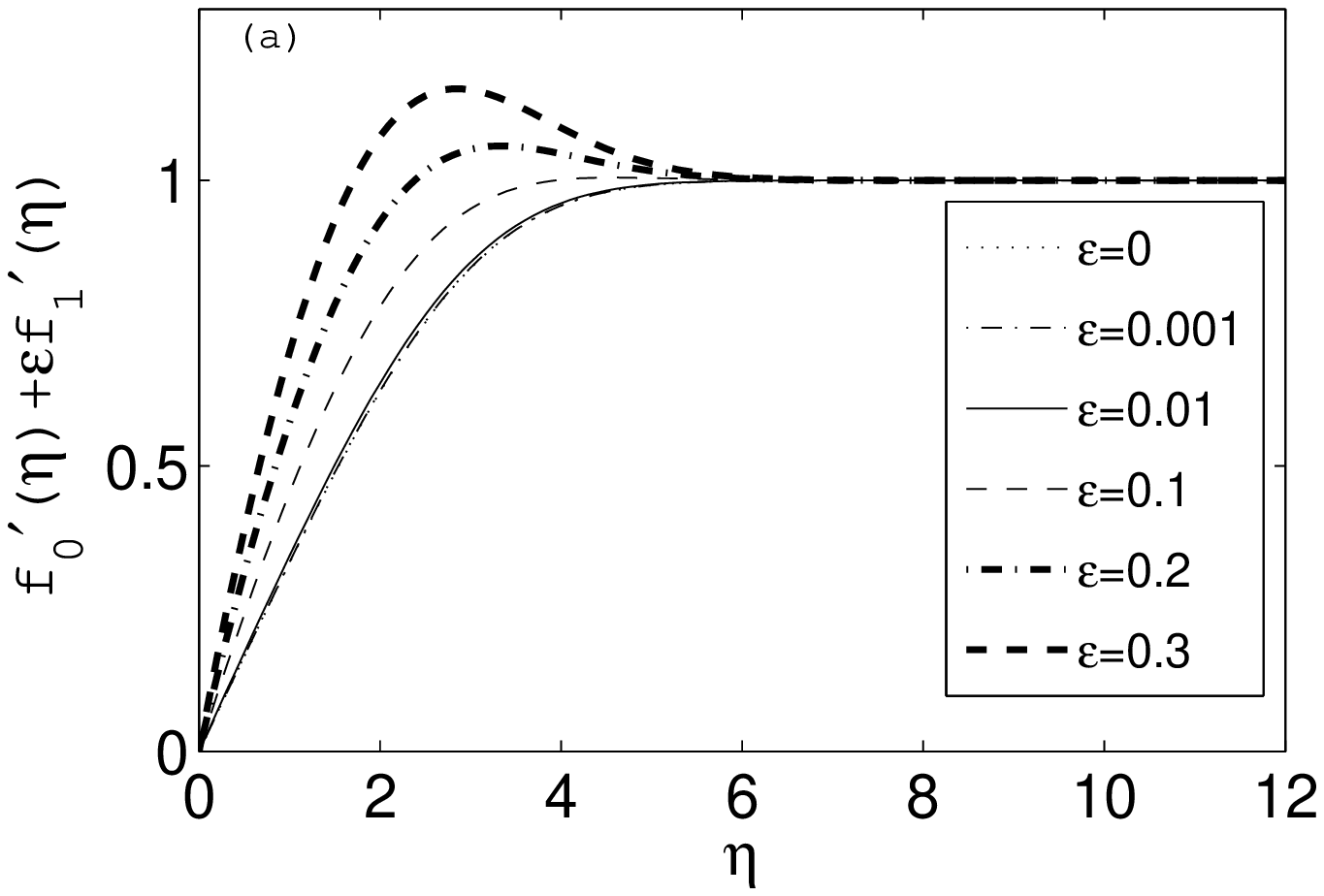}
\includegraphics[width=6.7cm]{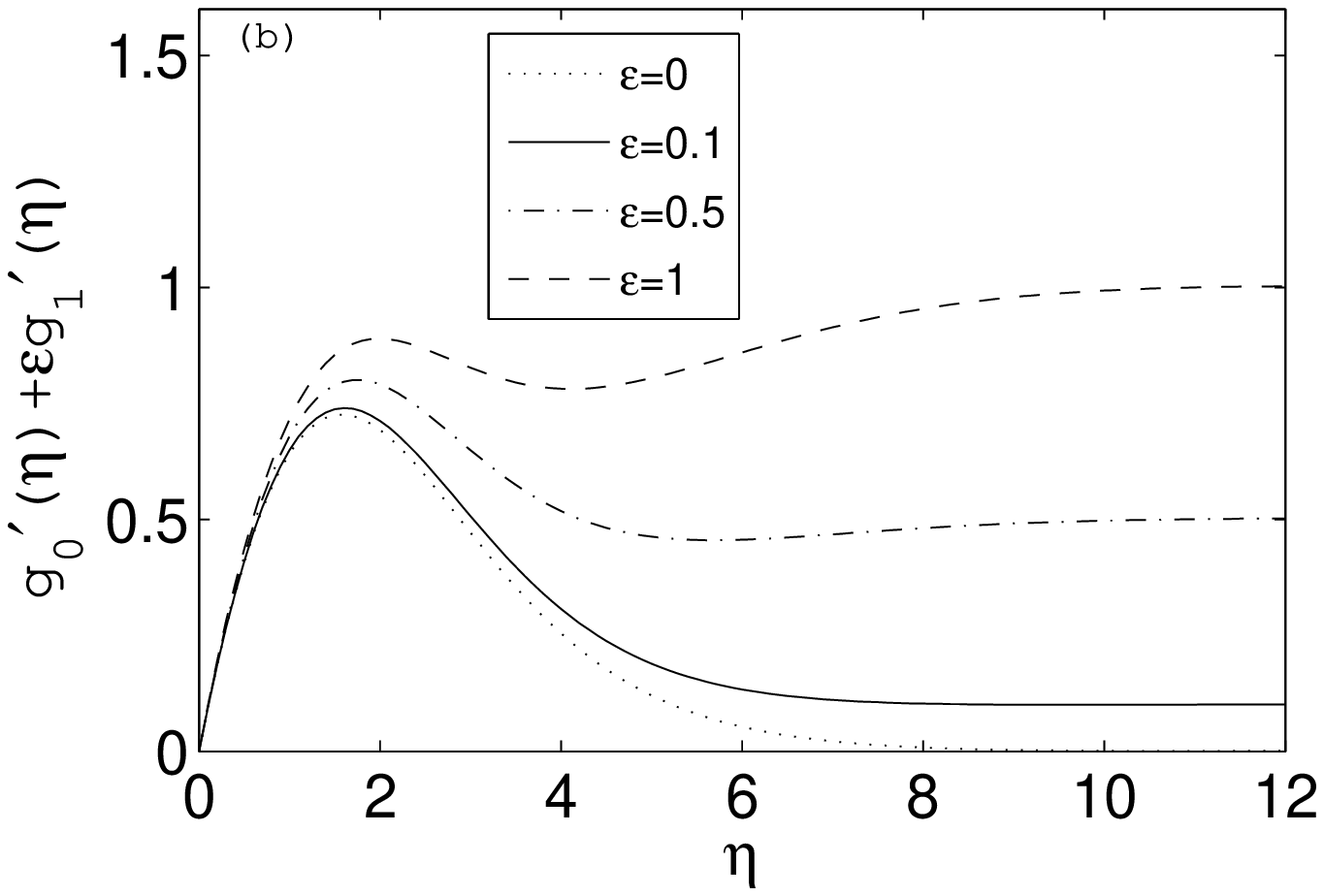}
\includegraphics[width=6.7cm]{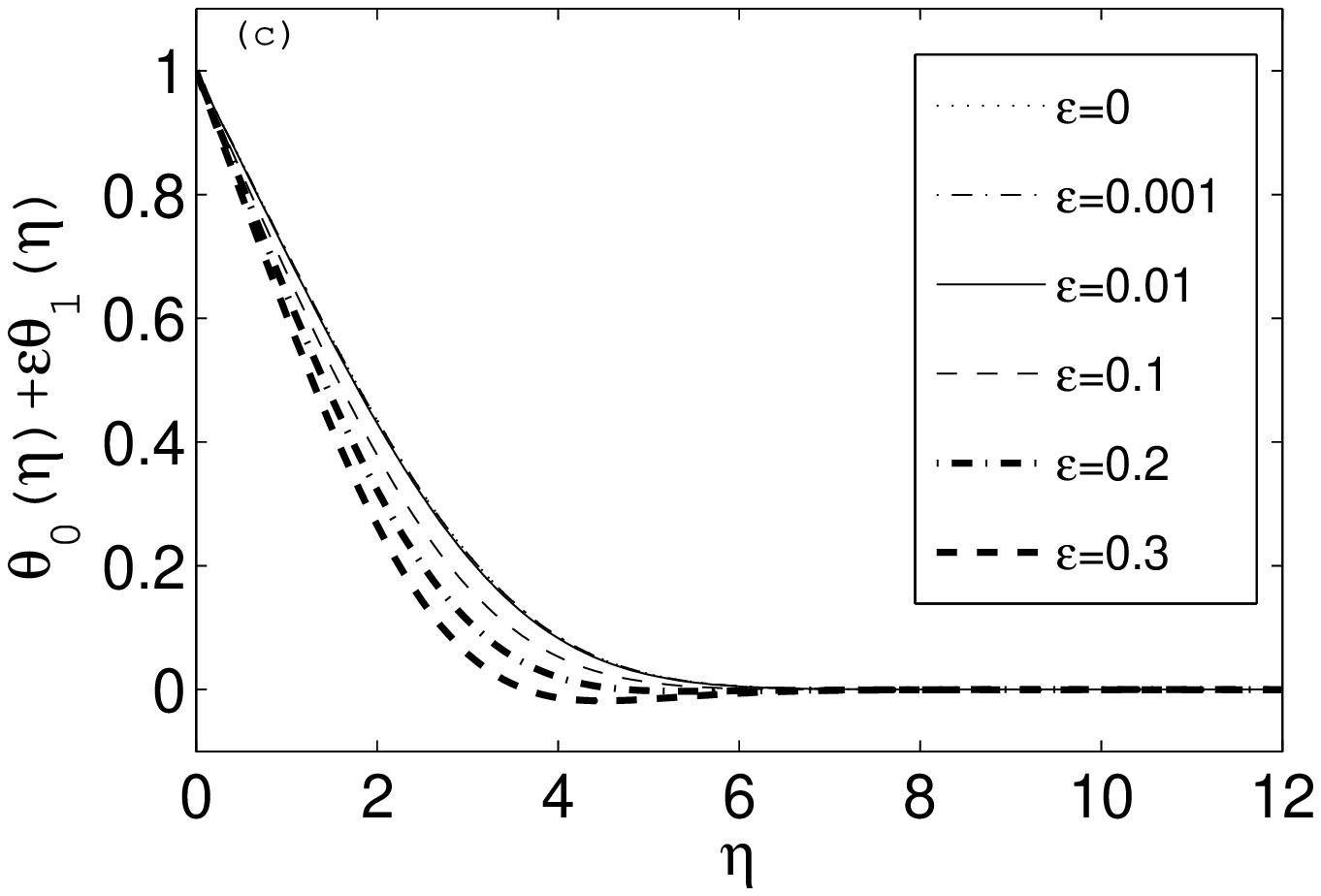}
\includegraphics[width=6.7cm]{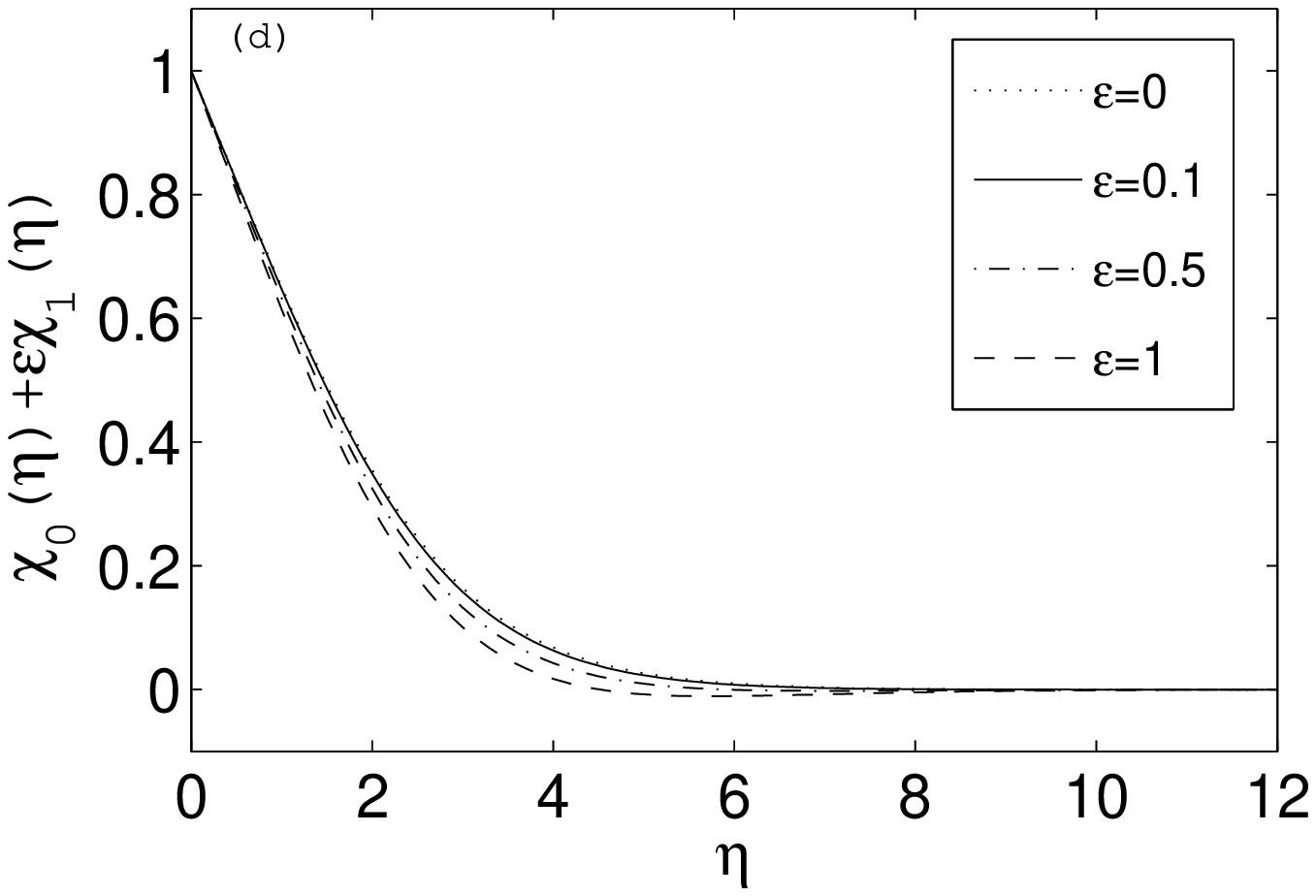}
\caption{Vertical profiles of the normalized downstream velocity and temperature resulting from 
a first-order perturbative expansion of forced (panels a, c) and natural convection (panels b, d). The values 
of the expansion parameter $\epsilon$ are indicated in the legend. The similarity variables are given by (\ref{Ap_1}). 
We also show examples for $\epsilon$ being unrealistically large in order to indicate the deviations better.} 
\label{fig7}
\end{center}
\end{figure}

Figure \ref{fig7} shows the resulting mean streamwise flow and temperature profiles 
for the case of $Pr=0.7$. The perturbation expansion has been carried out to the first order only 
and curves are plotted for different magnitudes of $\epsilon$ as given in (\ref{Ap_2}). Several 
aspects can be observed. The boundary layer flow is accelerated if buoyancy effects are added 
to the classical Blasius case as seen in panel (a) of the figure. The same holds if a purely natural convection layer is additionally 
driven by an outer flow such as the large-scale circulation in the 
present system (seen panel (b) of Fig. \ref{fig7}). The imposed outer flow causes a significant 
variation of the velocity profile. The modifications in the temperature are less pronounced. In both 
cases the resulting mean temperature profiles deviate slightly from the unperturbed results.   
\begin{figure}
\begin{center}
\includegraphics[width=14cm]{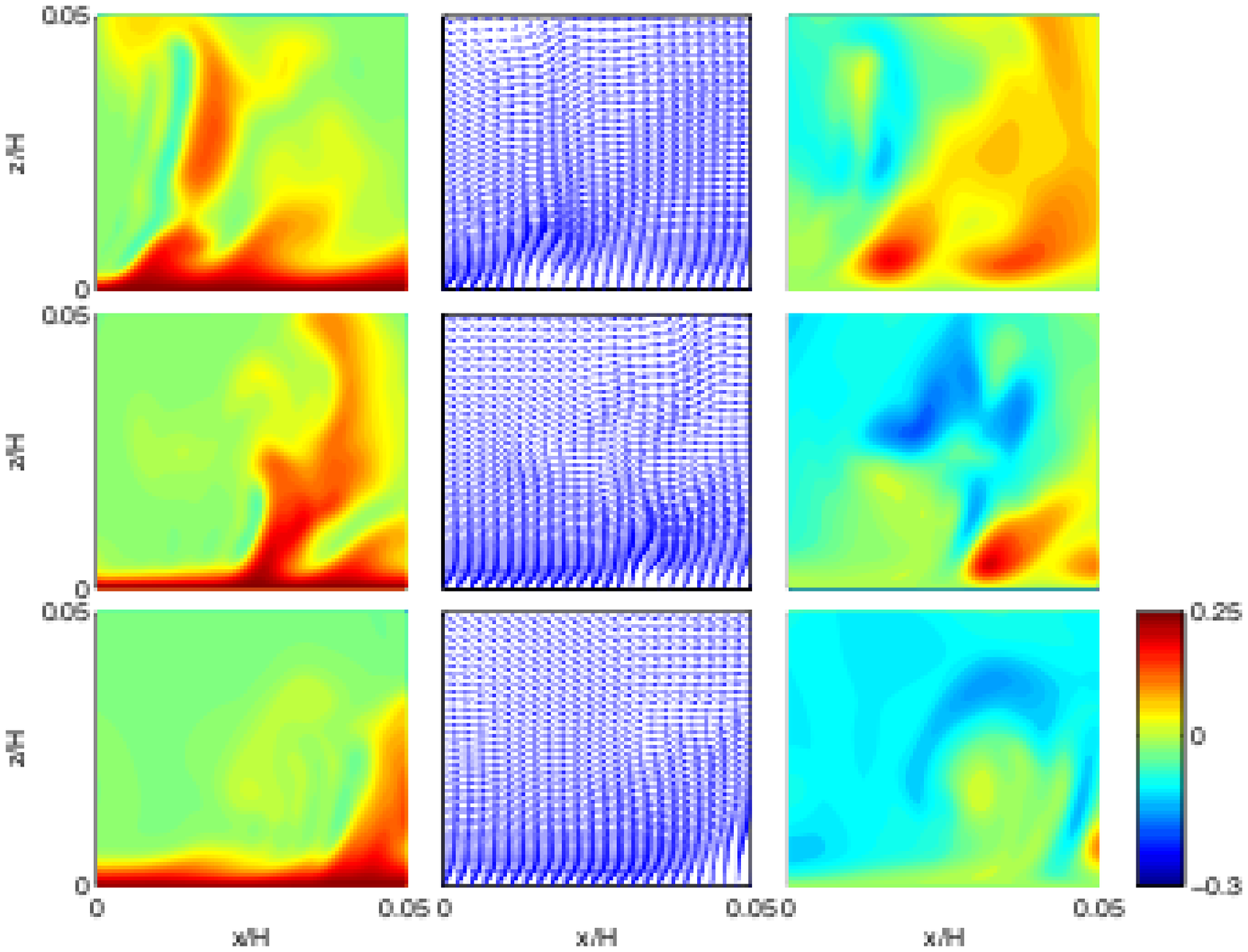}
\caption{(Color online) Sequence of three vertical plane cut snapshots to illustrate the detachment of a plume. 
The first column plots contours of the temperature $T$, the second is a vector plot of $(u_r, u_z)$, 
and the third plots contours of $u_{\phi}$. The pictures in the second and third rows lag behind
those of the first row by $0.2 T_f$ and $0.4 T_f$, respectively. The color legend (not shown here) for the temperature in the first column corresponds to an equidistant color scale
between zero (in blue) and one (in red). Scalar magnitudes for the azimuthal 
velocity component are indicated by the color bar (blue for negative and red for positive values ) in the right column. The three time instants 
are number 1, 5 and 9 out of a sequence of nine equidistant snapshots. Data are for $Ra=3\times 10^9$
where $\delta_T/H=0.0057$.} 
\label{sequence1}
\end{center}
\end{figure}
\begin{figure}
\begin{center}
\includegraphics[width=13cm]{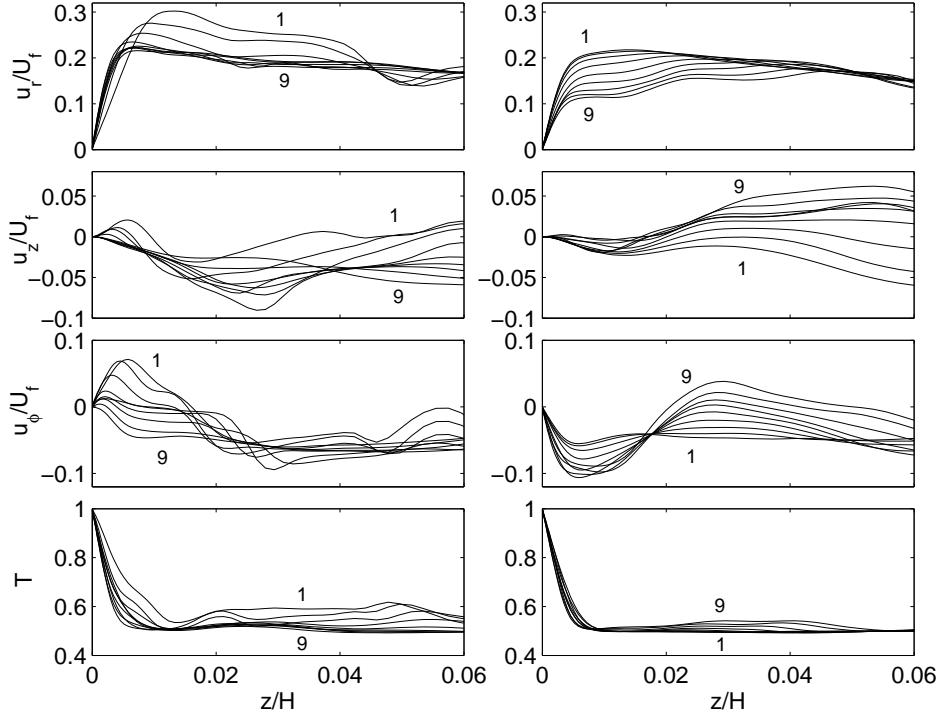}
\caption{Sequence of nine profiles that correspond with the data of Fig. \ref{sequence1} for the plume detachment 
phase (left column) and with the data of Fig. \ref{sequence2} for the post-plume phase (right column). They are 
obtained by an averaging in radial (or $x$-) direction over the window that is shown in Fig. \ref{sequence1}. From top to bottom:
Radial velocity component $u_r/U_f$,  vertical velocity component $u_z/U_f$, azimuthal velocity component 
$u_{\phi}/U_f$ and temperature $T$.} 
\label{sequence1a}
\end{center}
\end{figure}
\begin{figure}
\begin{center}
\includegraphics[width=14cm]{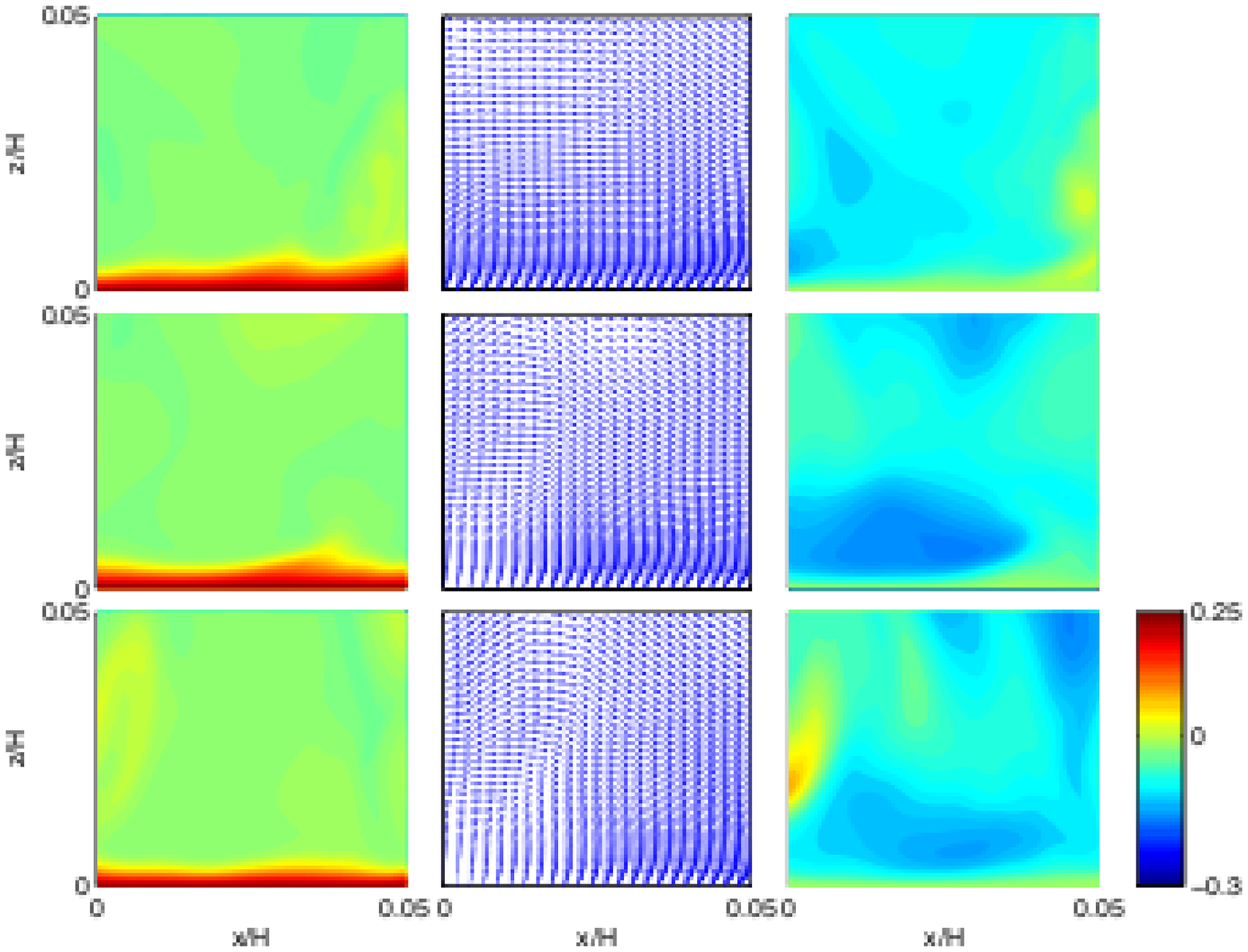}
\caption{(Color online) Sequence of three vertical plane cut snapshots to illustrate the phase after the detachment of a plume. The first column plots contours of the temperature $T$, 
the second is a vector plot of $(u_r, u_z)$, and the third plots contours of $u_{\phi}$. The pictures in the second and third rows lag behind
those of the first row by $0.2 T_f$ and $0.4 T_f$, respectively. The color legend (not shown here) for the temperature in the first column corresponds to an equidistant color scale
between zero (in blue) and one (in red). Scalar magnitudes for the azimuthal 
velocity component are indicated by the color bar (blue for negative and red for positive values ) in the right column. The three time instants are number 1, 5 and 9 out of a sequence of nine equidistant 
snapshots. Data are again for $Ra=3\times 10^9$.} 
\label{sequence2}
\end{center}
\end{figure}
\begin{figure}
\begin{center}
\includegraphics[width=14cm]{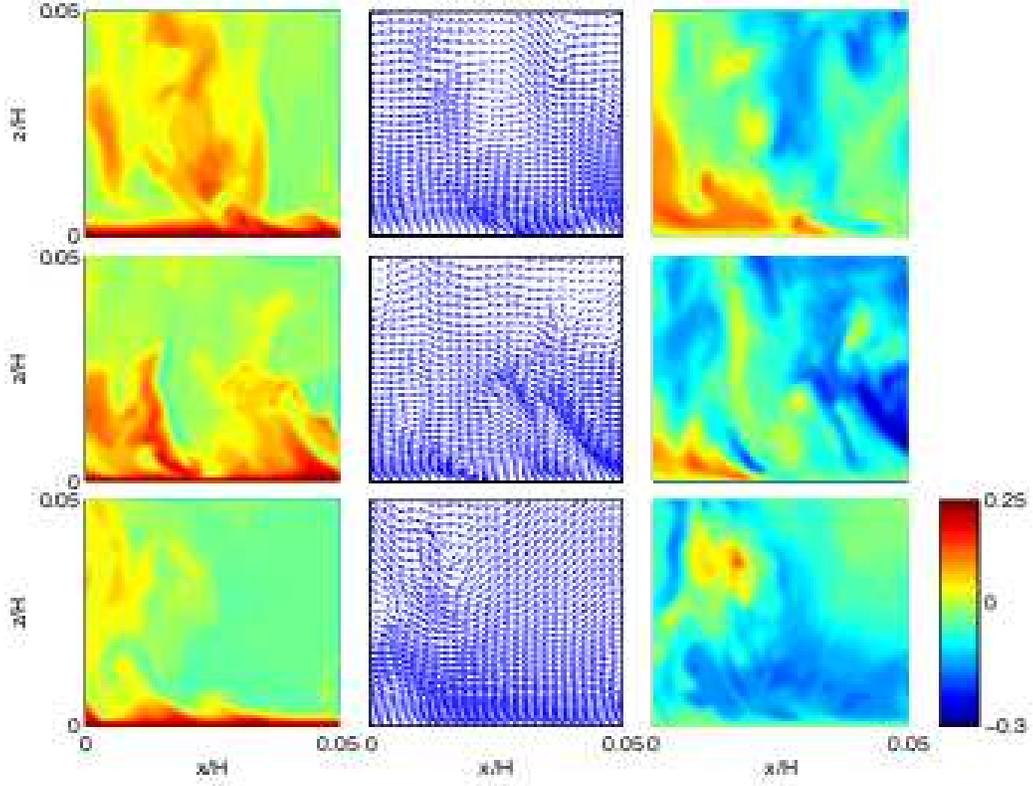}
\caption{(Color online) Same sequence of contour plots as in Fig. \ref{sequence1} for $Ra=3\times 10^{10}$ where $\delta_T/H=0.0026$.} 
\label{sequence1_10}
\end{center}
\end{figure}
\begin{figure}
\begin{center}
\includegraphics[width=13cm]{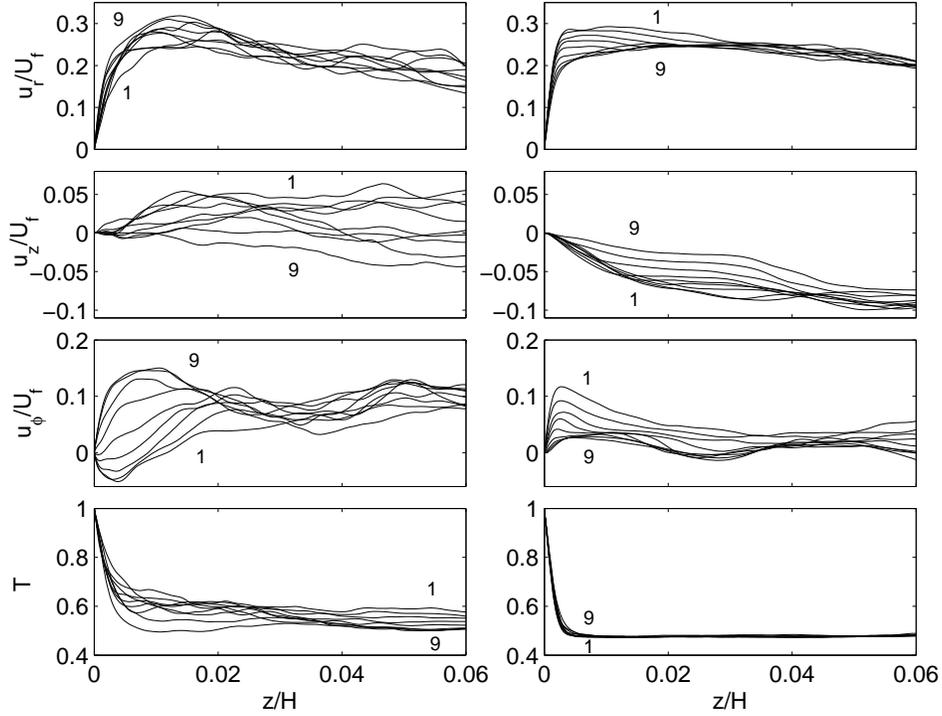}
\caption{Sequence of nine profiles that correspond with the data of Fig. \ref{sequence1_10} (left column) and Fig. \ref{sequence2_10} (right column).
All data are given in the same units as discussed in Fig. \ref{sequence1a}. } 
\label{sequence1a_10}
\end{center}
\end{figure}
\begin{figure}
\begin{center}
\includegraphics[width=14cm]{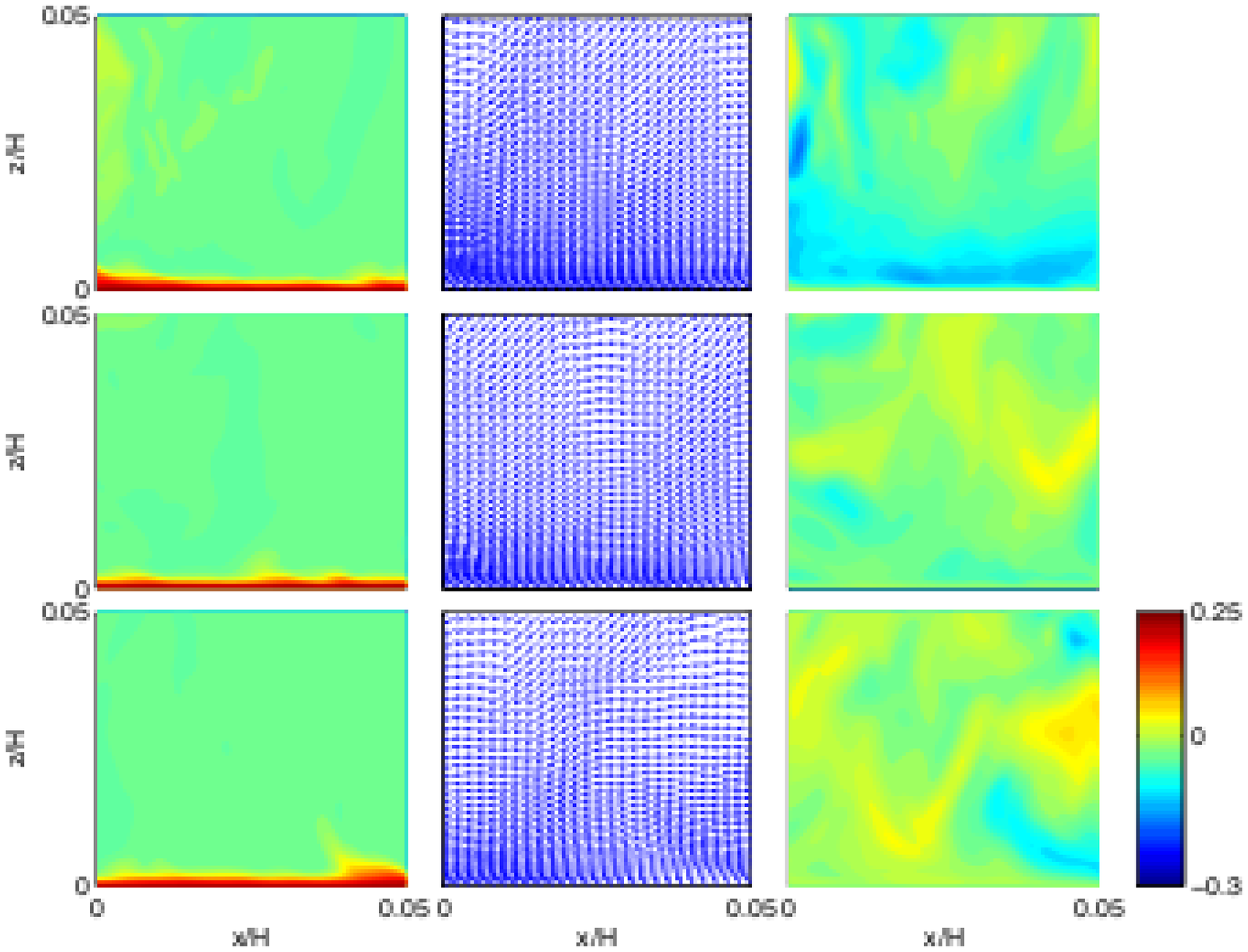}
\caption{(Color online) Same sequence of data as in Fig. \ref{sequence2} for $Ra=3\times 10^{10}$.} 
\label{sequence2_10}
\end{center}
\end{figure}

Let us now estimate the Rayleigh number dependence of the boundary layer thicknesses in both 
limiting cases of laminar boundary layers. The dependence of the thickness on the Reynolds 
number in forced convection is  given by (Schlichting, 1957) 
\begin{equation}
\delta_v\sim \frac{x}{Re_x^{1/2}}\,.
\end{equation}
By using a scaling relation between Reynolds and Rayleigh numbers for convection at $Pr\approx  1$ and $\Gamma=1$, which is taken from Ahlers et al. (2009), namely  $Re\sim Ra^{0.45}$, this  results in a Rayleigh number dependence $\delta_v\sim \delta_T \sim {Ra}^{-0.22}$ in the purely forced convection case. In a natural convection boundary layer,  the Grashof number is substituted 
with the Reynolds number and with the similarity variable  $\eta=z Gr_x^{1/5}/x$  one obtains
\begin{equation}
\delta_T\sim \frac{x}{Gr_x^{1/5}}\,.
\end{equation}
Again, if we are interested here in convection with Prandtl numbers around one such that  
$Gr_x\approx Ra$. It follows then that $\delta_v\sim\delta_T\sim {Ra}^{-0.2}$ which is very 
close to the forced case. Both scaling estimates suggest that the differences in the Rayleigh 
number dependence of the boundary layer thicknesses are rather small when both limits --
natural and forced convection -- are compared. With only two runs at different Rayleigh numbers at hand,
we are not able are to conduct scaling laws of the thicknesses with respect to $Ra$.  
  
\subsection{Boundary layer dynamics in a small observation window}
The present DNS give us the possibility to zoom into the boundary layer dynamics at higher Rayleigh numbers and to test how close the 
local profiles match with the results of the classical boundary layer theories that we just discussed. Out of the comprehensive data record, 
we have picked two characteristic dynamic sequences of the boundary layer structures -- a plume detachment event and the post-plume-detachment 
phase for which the boundary layer relaminarizes again. Each of these typical sequences covers a time lag of about $0.45 T_f$ for our data at 
both Rayleigh numbers. We consider  them as the two essential building blocks of the boundary layer dynamics. In order to make contact to 
the classical boundary layer theory, we analyze the fields again in a small vertical observation plane that is aligned with the instantaneous 
large-scale circulation. Our observation window has the size of length $\times$ height equal to $9 \delta_T\times 9\delta_T$ for $Ra=3\times 10^9$
and of $19 \delta_T\times 19\delta_T$ for $Ra=3\times 10^{10}$.  The dense temporal output of the data spans $35 T_f$ for $Ra=3\times 10^9$ and
$5 T_f$ for $Ra=3\times 10^{10}$ with a time interval of $0.05 T_f$ in both runs. 

A typical plume detachment event is seen in Fig. \ref{sequence1}  where the temperature is shown in the first, the velocity field projected into 
the plane in the second and the out-of-plane velocity component $u_{\phi}$ in the third column, respectively.  The rise of the hot fluid causes 
strong upward outflow that is connected with the plume detachment. This is in line with a strong inflow in the back of the plume due to the 
incompressibility of the flow. The whole detachment process is accompanied by a cross wind underlining the three-dimensionality of the whole dynamical process. The magnitude of the azimuthal velocity is comparable with the amplitudes 
of $V_{\perp,rms}$ in Fig. \ref{fig4}. Furthermore, the largest amplitudes of the azimuthal velocity component are found to be in line with the largest
values of $\delta_v(t)$ and $\delta_T(t)$. The plume detachment is thus one of the dynamical processes that cause the fluctuations of the boundary layer 
thicknesses. Our snapshot analysis also showed that the thickness variations are not significantly delayed with respect to each other which is in line with
the short lead time for $g(\tau)$ which we discussed in section \ref{s33}. The significant azimuthal velocity component confirms previous observations by 
Shishkina \& Wagner (2008) that a strong local vorticity vector field is aligned with line-like plume ridge.

The corresponding mean profiles of all velocity components and the temperature are shown in the left 
column of Fig. \ref{sequence1a}. They are obtained by averaging in the observational window with respect to the radial direction. The detachment 
is accompanied by a deceleration of the radial velocity and strong upward and downward flows into the bulk region as already described above. 
The temperature profiles deviate significantly  from the classical laminar boundary layer profile (see Fig. \ref{fig7}) as the hot fluid parcel leaves 
the observation area.   

The ambient post-plume-detachment phase is illustrated in Fig. \ref{sequence2}. At a first glance the flow and temperature fields seem to 
agree much better with the predictions from the laminar boundary layer theory. However, not too far away  from the wall still non-negligible upward and downward flows are present. The stratification of the temperature field is nearly unperturbed and the azimuthal component is more homogeneously distributed 
over the window in comparison to the detachment phase. This becomes also obvious from the plots in the right column of Fig. \ref{sequence1a} where 
the temperature profiles are much less perturbed than in  a plume detachment phase. Nevertheless, even in this phase the flow is three-dimensional
as we can see from the profiles of the azimuthal velocity component. Recall that the observational window in Figs. \ref{sequence1} and 
\ref{sequence2} has a height of $0.05 H$. Up to this distance from the wall, the maximum magnitude of the mean vertical velocity component is 
much smaller.    

The radial velocity and temperature profiles in both sequences indicate that the profiles vary strongly,
even over such a rather short dynamic sequence. The velocity is strongly enhanced in the boundary layer, 
as it is also resulting from the two-dimensional perturbative analysis, such as in the 
forced case (see Fig. \ref{fig7}(a)). Furthermore, the presented data indicate that the large-scale circulation is always strong enough such that the pure natural 
convection with a streamwise velocity that goes to 
zero, is not established (see Fig. \ref{fig7}(b)).      

We repeated this analysis for the second run at $Ra=3\times 10^{10}$. The qualitative picture remains unchanged for both phases, the 
plume detachment period and the post-plume phase. Note that the mean advection direction of the plumes is now opposite. The data are 
shown in the same way as for the lower Rayleigh number in Figs. \ref{sequence1_10} and \ref{sequence2_10}. As expected, the 
detaching plumes are more filamented and the boundary layer in the post-plume phase is thinner. The amplitude of the azimuthal velocity 
component remains significant as seen in Fig. \ref{sequence1a_10}. 

In both runs the profiles of $u_{\phi}$ show the following behaviour in the 
vicinity of the wall. In the plume detachment phase this velocity component changes the sign when moving forward in time from snapshot one to 
nine. This is not the case in post-plume phase. The differences between the temperature profiles for both phases are even more pronounced in
comparison to the lower Rayleigh number run.

The time lags of the plume detachment and post-plume phases have been calculated as follows. We take the radially averaged temperature 
field at $z\approx 5\delta_T$ for each snapshots in the window.  If this value exceeds the mean bulk temperature it is assigned with a detachment event 
otherwise it belongs to the post-plume phase. By applying this simple procedure, the time series is digitalized. The resulting step function has shorter and longer
time periods for both dynamical building blocks. The mean time of plume detachment and post-plume is about the same and gives about $0.45 T_f$ 
for $Ra=3\times 10^9$ where we had a sufficiently long time series. Combining both gives a typical cycle time of $T_f$ which is consistent with the
$2 T_{cross}$ from the fluctuating boundary layer thickness in section 3.4.
\begin{figure}
\begin{center}
\includegraphics[width=12cm]{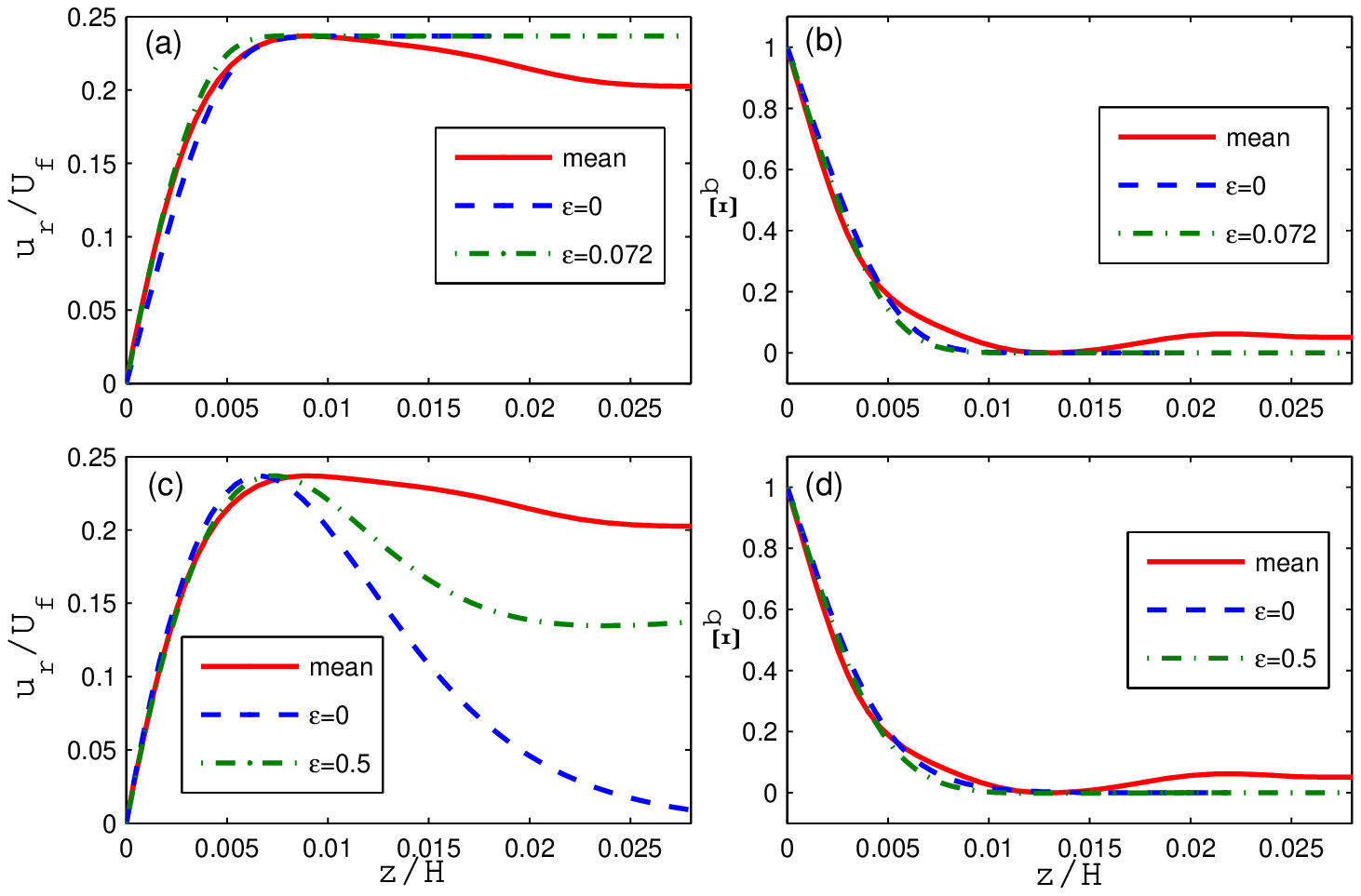}
\caption{(Color online) Matching of the time-averaged profiles of the plume detachment phase
with the predictions from boundary layer analysis. The profiles are obtained as 
time averages from Fig. \ref{sequence1a} (left column). Data are compared with 
purely forced and natural convection as well as with a corresponding first-order 
perturbative expansions. (a) Radial velocity in units of $U_f$ for forced convection. (b) 
Rescaled temperature for forced convection. (c) Radial velocity in units of $U_f$ for natural convection. (d) 
Rescaled temperature for natural convection. Data are 
the same as in Figs. \ref{sequence1} and \ref{sequence1a} (left column).} 
\label{sequence1b}
\end{center}
\end{figure}
\begin{figure}
\begin{center}
\includegraphics[width=12cm]{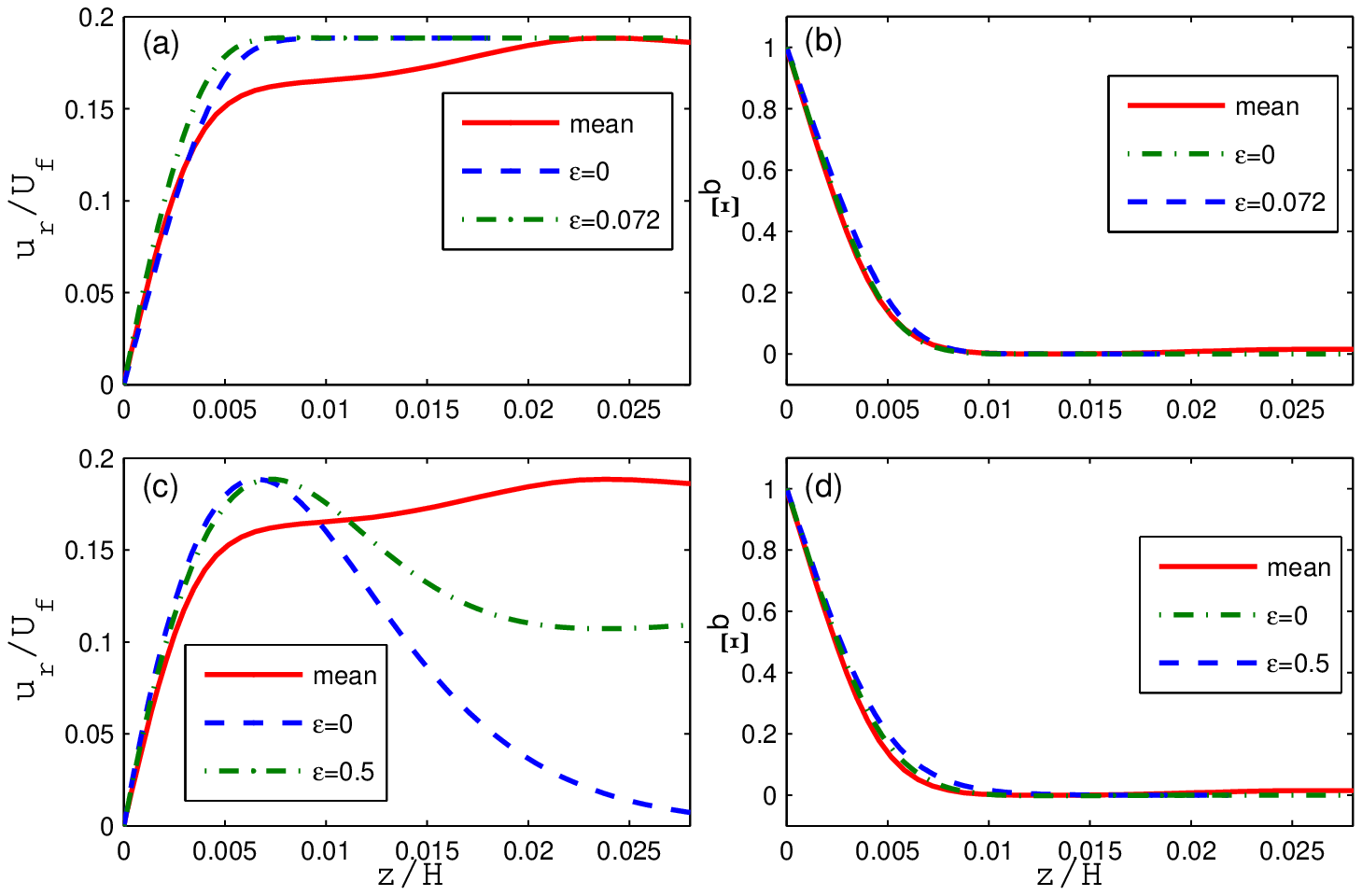}
\caption{(Color online) Matching of the time-averaged profiles of the post-plume-detachment phase
with the predictions from boundary layer analysis. The profiles are obtained as 
time averages from Fig. \ref{sequence1a} (right column). Panels (a)--(d) are as in Fig. 
\ref{sequence1b}. Data correspond to Figs. \ref{sequence1a} (right column) and \ref{sequence2}.}
\label{sequence2b}
\end{center}
\end{figure}
 
In Figs. \ref{sequence1b} and \ref{sequence2b} we try to match the time averaged profiles obtained
from the short dynamic sequences with the predictions from the mixed convection boundary layer theory including the 
first-order perturbation. Our profiles display again the features that we have detected in the original
time series analysis over much longer time intervals (see Fig. \ref{Radependence}). However we can now trace 
the slower increase of the temperature profile clearly back to the plume detachment events. 
Similar connection holds for the velocity profile in post-plume-detachment phase. The local dynamical 
behaviour suggests that the three-dimensional large-scale circulation is now connected to the boundary layer section.
Inflows from the top of our observation window are observed which cause large variations of the velocity 
profiles. These variations reach the same magnitude as in the plume detachment phase and manifest in the 
deviations for velocity profile $\langle u_r\rangle_r$ in the observation plane (see Fig. \ref{sequence2b}). We have thus shown that the simulation data combine elements of forced and natural convections. Neither in the plume detachment nor in the post-plume phase the theoretical profiles of both the temperature and velocity fields can be perfectly matched to the data. The dynamics close to the walls is always three-dimensional.
   
\section{Summary and outlook}  
We have studied the boundary layer dynamics of three-dimensional turbulent Rayleigh-B\'{e}nard 
convection in a cylindrical cell of aspect ratio one for Rayleigh numbers larger than $10^9$. Our studies 
are focussed on the convection in air with a $Pr=0.7$. The simulations provide access to the full 
spatial and temporal information in- and outside the thermal and velocity boundary layers. 

The large-scale circulation in the cell is varying in its direction and amplitude significantly providing a time-dependent 
driving of the boundary layer dynamics. The fluctuating LSC is in line with a strongly fluctuating thickness of both 
boundary layers which can be defined from instantaneous snapshots as suggested by Zhou \& Xia (2010). When 
these fluctuations are incorporated into a dynamical rescaling, the matching of the mean profiles to the Prandtl-Blasius-Pohlhausen 
theory improves. However, in the present cylindrical cell, deviations from the classical Prandtl-Blasius-Pohlhausen profiles will remain,
in particular for the temperature. 
The profiles do also not fit to the other limit case, natural convection. 

In the present DNS we aimed at connecting dynamical behavior in the boundary layer with the observed statistics. Our analysis found that the boundary layers 
follow a three-dimensional dynamics in all dynamical phases. This conclusion results from investigations of the pressure, the LSC and local
dynamic sequences. Pressure gradient components and temperature fluctuate strongly and follow non-Gaussian statistics. A significant flow perpendicular to a two-dimensional analysis plane is present during detachment, it is also observed in the post-detachment phase. These plumes form a line-like skeleton, but
are not found to be parallelly aligned.  Their detachment is accompanied with a significant variation of the boundary layer thicknesses
and a cross-wind (azimuthal velocity) with a significant amplitude.  

All these 
observational outcomes violate the assumptions made in deriving the similarity solutions in the classical boundary layer theories. Analyses in a point wise probe array as well as in a observational window support our findings. This limits also the applicability of two-dimensional plume models and causes to our point of view the deviations from both, the classical Prandtl-Blasius-Pohlhausen and the natural convection cases. It can be expected that the dynamics in the boundary layer will become increasingly intermittent when the 
Rayleigh number grows, a point that needs to be investigated further. Such an increasingly intermittent behavior would be typical for a transitional boundary layer which 
is ultimately  evolving towards a turbulent one at larger Rayleigh numbers. This interpretation would also be in line with the DNS results of the presently highest achievable Rayleigh numbers by Stevens et al. (2011). They found that the agreement of a dynamically rescaled thermal boundary layer with the Pohlhausen prediction worsens when $Ra$ grows.

One more point: the previous studies by Puthenveetil et al. (2011) as well as the recent experiments by Zhou and Xia (2010) suggest that the velocity 
boundary layer is much less perturbed when the Prandtl number is increased. In this case, the thermal 
boundary layer thickness becomes much smaller than the thickness of the velocity boundary layer. 
Plumes which detach will have a much narrower stem due to decreased thermal diffusion. We expect therefore that the agreement with results from the laminar boundary layer theory will improve.  This trend might however be compensated by an increasing number of fine-scale textures of the turbulent fields for increasing Rayleigh number. Our two streamline plots in Fig. 1 suggest this trend.  Further comprehensive numerical and experimental studies are thus necessary to answer these questions.     

\begin{acknowledgements}
The authors acknowledge support by the Deutsche Forschungsgemeinschaft (DFG) within the Research Group FOR1182, the Research Training 
Group GK1576 and the Heisenberg Program under Grant No. SCHU 1410/5-1. The largest DNS simulations have been carried out at the J\"ulich 
Supercomputing Centre (Germany)  under Grant No. HIL02 on one rack of the Blue Gene/P JUGENE. JS was supported in part by the National Science 
Foundation under Grant No. PHY05-51164 within the program ``The Nature of Turbulence'', held at the Kavli Institute of Theoretical Physics at 
the University of California in Santa Barbara. We thank Guenter Ahlers, Eberhard Bodenschatz, Ronald du Puits, Detlef Lohse, Baburaj Puthenveettil, 
Janet Scheel, Andr\'{e} Thess, Penger Tong, Roberto Verzicco, Sebastian Wagner and Ke-Qing Xia for fruitful discussions. Special thanks to Roberto Verzicco for  providing us his original code and to J. Rafael Pacheco for an improved Poisson solver, which saved us a significant amount of CPU time.  
\end{acknowledgements}

\section{Appendix: Perturbative expansion of the boundary layer equations}
We briefly review here the results reported in Sparrow \& Minkowycz (1962) and Stewartson (1959).
The similarity variable is given by 
\begin{equation}
\eta= \left\{
\begin{array}{ll}
z Re_x^{1/2}/x & 
             \;\;\;\text{for forced convection}\\
z Gr_x^{1/5}/x    &  
             \;\;\;\text{for natural convection}
\end{array} \right.
\label{Ap_1}
\end{equation}
and the expansion parameter is given by
\begin{equation}
\epsilon= \left\{
\begin{array}{ll}
Gr_x/Re_x^{5/2} & 
             \;\;\;\text{for forced convection}\\
Re_x/Gr_x^{2/5}    &  
             \;\;\;\text{for natural convection}
\end{array} \right.
\label{Ap_2}
\end{equation}
Since the problem at hand is two-dimensional, one uses the stream function instead of the velocity components which automatically satisfies the incompressibility condition (\ref{bleq3}). In the {\em forced} convection case the following expansions are taken
\begin{eqnarray}
\psi(x,z)&=&\sqrt{\nu x V_{\infty}} \left[\sum_{m=0}^{\infty}\epsilon^m f_m(\eta)\right]\,,\\
T(x,z)&=&T_{\infty}+(T_w-T_{\infty})\left[\sum_{m=0}^{\infty} \epsilon^m \theta_m(\eta)\right]\,,
\end{eqnarray}
resulting for example in the following expressions for the velocity components
\begin{eqnarray}
u_x(x,z)&=&V_{\infty} \left[\sum_{m=0}^{\infty}\epsilon^m f^{\prime}_m(\eta)\right]\,,\\
u_z(x,z)&=&\frac{V_{\infty}}{2\sqrt{Re_x}}
 \left[(\eta f^{\prime}_0(\eta)-f_0(\eta))+\sum_{m=1}^{\infty}\epsilon^m \left(\eta f^{\prime}_m(\eta)+
 \left(\frac{m}{2}-1\right)
 f_m(\eta)\right)\right]\,,
\end{eqnarray}
where primes denote derivatives with respect to $\eta$.
The expansion generates in order $\epsilon^0$ the classical Prandtl-Blasius-Pohlhausen equations 
\begin{eqnarray}
\label{bleq1b}
f_0^{\prime\prime\prime}+\frac{1}{2}f_0^{\prime\prime}f_0&=&0\,,\\
\label{bleq2b}
\theta_0^{\prime\prime}+\frac{Pr}{2}f_0\theta_0^{\prime}&=&0\,.
\end{eqnarray}
The boundary conditions are $f_0(0)=f_0^{\prime}(0)=0$, $\theta_0(0)=1$ and 
$f_0^{\prime}(\infty)=1$, $\theta_0(\infty)=0$. The order $\epsilon^1$ reads then
\begin{eqnarray}
\label{bleq1c}
f_1^{\prime\prime\prime}+f_0^{\prime\prime}f_1+\frac{1}{2}f_0 f_1^{\prime\prime}
-\frac{1}{2}f_0^{\prime} f_1^{\prime}-\frac{1}{2}h_0+\frac{\eta}{2}h^{\prime}_0&=&0\,,\\
\label{bleq2c}
h_0^{\prime}&=&\theta_0\,,\\
\label{bleq3c}
\theta_1^{\prime\prime}+\frac{Pr}{2}f_0\theta_1^{\prime}-\frac{Pr}{2} f_0^{\prime}\theta_1
+Pr\,\theta_0^{\prime}f_1&=&0\,.
\end{eqnarray}
The additional boundary conditions follow to $f_1(0)=f^{\prime}_1(0)=\theta_1(0)=0$ and 
$f_1^{\prime}(\infty)=\theta_1(\infty)=h_0(\infty)=0$. The last two terms of (\ref{bleq1c}) containing 
$h_0$ and $h_0^{\prime}$ as well as Eq. (\ref{bleq2c}) arise from the pressure term.
In {\em natural} convection, the expansions are adapted to 
\begin{eqnarray}
\psi(x,z)&=&\sqrt[5]{\nu^3 g \alpha (T_w-T_{\infty}) x^3} \left[\sum_{m=0}^{\infty}\epsilon^m g_m(\eta)\right]\,,\\
T(x,z)&=&T_{\infty}+(T_w-T_{\infty})\left[\sum_{m=0}^{\infty} \epsilon^m \chi_m(\eta)\right]\,.
\end{eqnarray}
The order $\epsilon^0$ was first discussed by Stewartson (1958) and given by 
\begin{eqnarray}
\label{bleq1e}
g_0^{\prime\prime\prime}+\frac{3}{5}g_0^{\prime\prime}g_0-\frac{1}{5}g^{\prime}_0g^{\prime}_0
-\frac{2}{5}k_0+\frac{2}{5}\eta k^{\prime}_0&=&0\,,\\
\label{bleq2e}
k_0^{\prime}&=&\chi_0\,,\\
\label{bleq3e}
\chi_0^{\prime\prime}+\frac{3Pr}{5}g_0\chi_0^{\prime}&=&0\,.
\end{eqnarray}
The boundary conditions are $g_0(0)=g_0^{\prime}(0)=0$, $\chi_0(0)=1$ and 
$g_0^{\prime}(\infty)=\chi_0(\infty)=k_0(\infty)=0$. The perturbative expansion to mixed convection 
with order $\epsilon^1$ reads 
\begin{eqnarray}
\label{bleq1f}
g_1^{\prime\prime\prime}+\frac{3}{5}g_1^{\prime\prime}g_0-\frac{1}{5}g^{\prime}_1 g^{\prime}_0
+\frac{2}{5} g^{\prime\prime}_0 g_1
-\frac{1}{5}k_1+\frac{2}{5}\eta k^{\prime}_1&=&0\,,\\
\label{bleq2f}
k_1^{\prime}&=&\chi_1\,,\\
\label{bleq3f}
\chi_1^{\prime\prime}+\frac{3Pr}{5}g_0\chi_1^{\prime}+\frac{Pr}{5}g_0^{\prime}\chi_1
+\frac{2Pr}{5}\chi_0^{\prime} g_1&=&0\,,
\end{eqnarray}
with $g_1(0)=g^{\prime}_1(0)=\chi_1(0)=\chi_1(\infty)=k_1(\infty)=0$ and $g^{\prime}_1(\infty)=1$. 
Again, $k_0$ and $k_1$ arise from the pressure term. Equations (\ref{bleq1b})--(\ref{bleq3c}) and 
(\ref{bleq1e})--(\ref{bleq3f}) have been solved in order to obtain the results displayed in Fig. \ref{fig7}.

\end{document}